\documentclass[a4paper,journal]{IEEEtran}

\usepackage{epsfig}
 \usepackage[cmex10]{amsmath}
 \usepackage{mathrsfs,amssymb}

\usepackage{array}

\usepackage{mdwmath}
\usepackage{mdwtab}
\usepackage[active]{srcltx}

\def\map#1{\mathscr{#1}}
\def\Tr{\mathrm{Tr}}

\def\Rng{\mathrm{Rng}}
\def\Lin{{\mathcal L}}
\def\>{\rangle}
\def\<{\langle}
\def\bb{\<\!\<}
\def\kk{\>\!\>}
\def\vect#1{\bf #1}
\def\n#1{\left|\!\left|#1\right|\!\right|}

\def\sH{\mathcal H}
\def\sK{\mathcal K}

\def\d{\operatorname d}

\newcommand{\Ket}[1]{| #1 \rangle \! \rangle}
\newcommand{\Bra}[1]{\langle \! \langle #1 |}

\newcommand{\KetBra}[2]{\Ket{#1} \Bra{#2}}

\begin{document}
%
\title{Optimal quantum tomography}
%
%
%

\author{Alessandro Bisio, Giulio~Chiribella, 
  Giacomo~M.~D'Ariano, Stefano Facchini, 
  and~Paolo~Perinotti}
\maketitle

\begin{abstract}
  The present short review article illustrates the latest theoretical
  developments on quantum tomography, regarding general optimization
  methods for both data-processing and setup.  The basic theoretical
  tool is the {\em informationally complete measurement}. The
  optimization theory for the setup is based on the new theoretical
  approach of {\em quantum combs}.
\end{abstract}

\begin{IEEEkeywords}
  Quantum Tomography, Quantum Process Tomography, Quantum Information
\end{IEEEkeywords}

%
\IEEEpeerreviewmaketitle

\section{Introduction}

\IEEEPARstart{F}{ine} calibration of apparatuses is the basis of any precise experiment, and the
quest for precision and reliability is relentlessly increasing with the strict requirements of the
new photonics, nanotechnology, and the new world of quantum information. The latter, in particular,
depends crucially on the reliability of processes, sources and detectors, and on precise knowledge
of all sources of noise, e.g. for error correction. 

But what does it mean to calibrate a quantum device? It is really a much harder task than
calibrating a classical ``scale''. For example, for calibrating a photo-counter, we don't have
standard sources with precise numbers of photons---the equivalent of the ``standard weights'' for
the scale. Even worst, we never know for sure that all photons have been actually absorbed by the
detector. The practical problem is then to perform a kind of {\em quantum calibration} to determine
in a purely experimental manner (by relying on some well established measuring instruments) the
quantum description of our device, without the need of a detailed theoretical knowledge of its inner
functioning---being it a measuring apparatus, a quantum channel, a quantum gate, or a source of
quantum states.

And here it comes the powerful technique of quantum tomography.
Originally invented for determining the quantum state of radiation
(for recent reviews see the book \cite{libroparis} and e.~g. Refs.
\cite{review_cerf,revtomo}, it soon became the universal measuring
technique by which one can determine any ensemble average and measure
the fine details of any quantum operation, channel, or measuring
instruments---objects that before were just theoretical tools (for
history and references see next section).

In the present short review article we will illustrate our latest theoretical developments on
quantum tomography, consisting in a first systematic theoretical approach to optimization of both
data-processing and setup. Therefore, apart from the historical excursus of the next section, where
we mention the relevant contributions from other authors, the body of the paper is focus only on our
theoretical work.  

The basic tool of the theoretical approach is the {\em informationally complete measurement}
\cite{bush} (see Refs. \cite{udet,infoc} for applications in the present context)---corresponding to
the mathematical theory of operator bases. The optimization of data-processing \cite{darper} relies
on the fact that as an operator basis the informationally complete measurement is typically linearly
dependent, allowing different expansion coefficients, which can be then optimized, according to
specific criteria.  The optimization theory for the setup \cite{opttomo}, on the other hand, needs
the new theory of quantum combs and quantum testers \cite{qca}, novel powerful notions in quantum
mechanics, which generalize those of quantum channel and of POVM (positive-operator-valued measure).
These will be briefly reviewed in the section before conclusions. As the reader will see, the
theoretical framework is sufficiently general and mature for a concrete optimization in the lab,
i.e. accounting for realistic bounded resources, and this will be the direction of future
development of the field.

\section{Historical excursus\label{s:excursus}}

Quantum tomography is a relatively recent discipline. However, the
possibility of ``measuring the quantum state'' has puzzled physicists
in the last half century, since the earlier theoretical studies of
Fano \cite{fano} (see also Pauli in Ref. \cite{pauli}).  That more
than two observables---actually a complete set of them, a so-called
{\em quorum of observables} \cite{bandpark,despag}---are needed for a
complete determination of the density matrix was immediately clear
\cite{fano}.  However, in those years it was hard to devise concretely
measurable observables other than position, momentum and energy (Royer
pointed out that instead of measuring varying observables one can vary
the state itself in a controlled way, and measure e.~g.  just its
energy \cite{Royer}). For this reason, the fundamental problem of
determining the quantum state remained at the level of mere
speculation for many years. The issue finally entered the realm of
experiments only less than twenty years ago, after the pioneering
experiments by Raymer's group \cite{reim}, in the domain of quantum
optics. Why quantum optics?  Because in quantum optics, differently
particle physics, there is the unique opportunity of measuring all
possible linear combinations of position and momentum of a harmonic
oscillator, representing a single mode of the electromagnetic field.
Such measurement can be achieved by means of a balanced homodyne
detector, which measures the quadrature $X_\phi=\frac{1}{2}
\left(a^{\dagger} e^{i\phi}+a e^{-i\phi}\right)$ of a field mode at
any desired phase $\phi$ with respect to the local oscillator (LO) [as
usual $a$ denotes the annihilator of the field mode]. The first
technique to reconstruct the density matrix from homodyne
measurements---so called {\em homodyne tomography}---originated from
the observation by Vogel and Risken \cite{vrisk} that the collection
of probability distributions $\{p(x,\phi)\}$ for $\phi\in[0,\pi)$ is
just the Radon transform---i.e.  the {\em tomography}---of the Wigner
function $W$. Therefore, by a Radon transform inversion, one can
obtain $W$, and from $W$ the matrix elements of the density operator
$\rho$. This first method, however, works fine only for high number of
photons or for {\em almost classical} states, whereas in the truly
quantum regime is affected by the smoothing needed for the Radon
transform inversion. The main physical tool, however---i.e.  using
homodyning---was a perfectly good idea: one just needed to process the
experimental data properly.
\par In Ref.\cite{dar} the first exact technique was given for measuring experimentally the matrix
elements of $\rho$ in the photon-number representation, by just averaging functions of homodyne
data. After that, the method was further simplified \cite{ulf2}, and the feasibility for nonunit
quantum efficiency $\eta<1$ at detectors---above some bounds---was established. Further improvements
in the numerical algorithms made the method so simple and fast that it could be implemented easily on
small PCs, and the method became quite popular in the laboratories (for the earlier progresses and
improvements the reader can see the old review \cite{bilktomo}). In the meanwhile there has been an
explosion of interest on the subject of {\em measuring quantum states}, with hundreds of papers,
both theoretical and experimental.  The exact homodyne method has been implemented experimentally to
measure the photon statistics of a semiconductor laser \cite{raymer95}, and the density matrix of a
squeezed vacuum\cite{exper-onemode2}.  The success of optical homodyne tomography has then
stimulated the development of state reconstruction procedures for atomic beams \cite{atobeams}, the
experimental determination of the vibrational state of a molecule \cite{diatomolecu}, of an ensemble
of helium atoms \cite{exper-freeatoms}, and of a single ion in a Paul trap \cite{leibfried}, and
different state reconstruction methods have been proposed (for an extensive list of references of
these first pioneering years, see e.g. Ref. \cite{twoself}).
\par Later the method of quantum homodyne tomography has been generalized to the estimation of an
arbitrary observable of the field \cite{tokyo}, with any number of modes \cite{1LO}, and, to
arbitrary quantum systems via group theory \cite{chicago,tomo_group,tomo_genova}, and with a general
method for unbiasing noise \cite{chicago,tomo_group}.  Eventually, it was recognized that the
general data-processing is just an application of the theory of operator
expansions\cite{gentomo_ortho,gentomo_quorum}, which lead to identify quantum tomography as an {\em
  informationally complete} measurement \cite{prugo}---a generalization of the concept of {\em
  quorum of observables} \cite{bandpark,despag}.\par 

State reconstruction was extended to the case where an
\emph{incomplete} measurement is performed. In this the reconstruction
of the full 
 density matrix of the system is actually impossible, and
one can only estimate the
 state that best fits the measured data
applying the Jaynes's maximum entropy principle (\emph{MaxEnt}) \cite
{buz}.
 When one has some non-trivial prior information the fit can be
improved by minimizing the Kullback-Leibler
 distance from a given
state which represents this \emph{a priori} information \cite{olipa}.\par

At the same time, in alternative to the averaging
data-processing strategy of the original method
 \cite{dar}, it was
recognized in Refs. \cite{hrad,banaszek} the possibility of
implementing a
 maximum likelihood strategy for reconstructing the
diagonal of the density matrix, and later for the
 full matrix
\cite{maxlik}. An advantage of the maximum likelihood strategy is that
the density
 matrix is constrained to be positive, whereas positivity
can be violated in the fluctuations of the
 averaging strategy. In
addition, the maximum likelihood often allows to reduce dramatically
the
 number of experimental data for achieving the same statistical
error, at the expense of a bias,
 which is however negligible in many
cases of practical interest.  However, there is a drawback: this
 is
the need of estimating the full density matrix (the strategy is
essentially a maximization of the
 joint probability of the full
data-set over all possible density matrices, or Bayesian variations
of
 such maximization accounting for prior knowledge\cite{b-k}).
This, on one side requires a cutoff of
 the dimension of the Hilbert
space when infinite (such as for the harmonic oscillator, as in
homodyne tomography), thus introducing the mentioned bias; on the
other side it has computational
 and memory complexities which
increase exponentially with the number of systems for a joint
tomography on multiple systems. On the contrary, the averaging
strategy for any desired expectation
 value needs just to average a
single function of the experimental outcome, without needing the full
matrix, and this includes as a special case the evaluation of single
matrix element itself, whence
 without necessitating a dimensional
cutoff.

 Contemporary to this preliminary evolution of
data-processing methods, there has been also a
 parallel evolution in
the tomographic setup design. It was realized that it is possible not
only
 states, but also channels \cite{nielchu,poya}---the so-called
(standard quantum) {\em process
 tomography} (SQPT)---based on the
idea of tomographing the outputs of a channel corresponding to a
 set
of input states making an operator basis for all density matrices.
However, soon later it was
 recognized (first for the diagonal matrix
elements in the number basis of an optical process
\cite{tomo_hamilt}, then in general for any channel
\cite{darlop,leung} that the same process
 tomography can be actually
achieved using just a single input state entangled (with maximal
Schmidt
 number) with an ancilla---the so-called {\em ancilla-assisted
  process tomography}
 (AAPT)---exploiting the ``quantum parallelism''
of the entangled input state which plays the role of
 a
``superposition of all possible input states''. This can have a great
experimental advantage when
 the basis of states is not easily
achievable experimentally, whereas the entangled state is, as in
 the
case of homodyne tomography where it is easy to achieve such entangled
state from parametric
 down-conversion of vacuum, whereas it is hard
to achieve photon-number states (see however, Ref.
 \cite
{sacchialpha}, where a set of random coherent states have been
proposed as a basis). As later
 proved in Ref.  \cite{faith}, and
experimentally verified in Refs.  \cite{altep}, almost any joint
system-ancilla state can be exploited for AAPT.  On the other hand,
the same AAPT has been extended
 to quantum operations and to
measuring apparatus \cite{calipovm,tomo-rew-spring-lopresti} (former
theoretical proposals for calibration of detectors were published
without ancilla
 \cite{fiurahrad,fiura}, and even ancilla-assisted
\cite{sanso}). Later, by another kind of quantum
 parallelism, it was
recognized that one can also estimate the ensemble average of all
operators of a
 quantum system by measuring only one fixed "universal"
entangled observable on an extended Hilbert
 space
\cite{tomo_holography}---a truly {\em universal observable}.  At this
point the tomographic
 method had reached the stage in which a single
fixed apparatus (single preparation of the input and
 single
observable at the output) is needed, in principle reducing enormously
the experimental
 complexity for joint tomography on many systems
(complexity 1 versus exponential complexity).
 
 After the first
experimental SQPT by NMR \cite{ccl}, AAPT was experimentally proved in
Refs.
 \cite{dema,altep}, for photon polarization qubit quantum
operations, exploiting spontaneous
 parametric downconversion in a non
linear crystal as a source of entangled states.
 
 As in the case of
state tomography, the freedom in the choice of the experimental
configuration
 poses the natural question of what is the optimal setup
for a given figure of merit. In Ref.
 \cite{mohlid} the issue of
minimizing the number of different experimental configurations needed
for
 process tomography was raised again, and a the so-called Direct
Characterization of Quantum Dynamics
 (DCQD) setup was introduced for
qubits, later generalized to arbitrary finite dimensional systems
\cite{mohlid2}. The proposed protocol starts from the expression of
the Choi-Jamio\l kowski operator
 (also called $\chi$-matrix)
$\map{C}(\rho)=\sum_{mn}\chi_{mn}A_m \rho A_n^\dag$ of the quantum
channel $\map{C}$ operating on the inputs state $\rho$, choosing for
the basis $\{A_m\}$ the
 shift-and-multiply group elements, and then
uses techniques from error detection for the estimation
 of parameters
$\chi_{mn}$ from estimated error probabilities. The DCQD approach is
interesting
 because of the interpretation of Process Tomography in
terms of error detection, however, it does
 not provide any optimality
argument in terms of number of experimental configurations, apart from
a
 vague resource analysis \cite{mohlid3}. A similar scheme was
introduced in Ref.  \cite{paz}, where
 the authors provide a method
for process tomography that allows to separately reconstruct the Choi
operator matrix elements in a fixed basis based by Haar-distributed
input state sampling. The
 authors exploit spherical $2$-designs
\cite{delsarte} in order to discretize the required averaging
 over
the group $\mathbb{SU}(d)$.
 
 In more recent years some experiments
in the continuous-variable domain were performed both for process
tomography \cite{lvov} and for measurement calibration \cite{plenio},
however both experiments
 exploited the SQPT technique, while no AAPT
experiments with continuous variable systems have been
 reported so
far. Many tomographic experiments on different kinds of quantum
systems have been
 performed, like atoms in optical lattices
\cite{opla}, cold ions in Paul traps \cite{blatt}, NMR
 probed
molecules \cite{nmr}, solid state qubits \cite{solstate}, and quantum
optic cavity modes
 interacting with atoms \cite{cv}.
 
 In the last
decade the interest in quantum tomography grew very fast with the
increasing number of
 applications in the hot field of quantum
information, allowing testing the accuracy of
 state-preparation and
calibration of quantum gates and measuring apparatuses. One should
realize
 that the whole technology of quantum information crucially
depends on the reliability of processes,
 source and detectors, and on
precise knowledge of sources of noise and errors. For example, all
error correction techniques are based on the knowledge of the noise
model, which is a prerequisite
 for an effective design of correcting
codes \cite{shor,steane,knillaf}, and Quantum Process
 Tomography
allows a reliable reconstruction of the noise and its decoherence free
subspaces without
 recurring to prior assumptions on the noisy
channels \cite{mitch}. The increasingly high confidence
 in the
tomographic technique, with the largest imaginable flexibility of
data-processing, and
 expanding outside the optical domain in the
whole physical domain, grew the appetite of
 experimentalists and
theoreticians posing increasingly challenging problems. The relevant
issues
 were now to establish the optimal tomographic setups and
data-processing, and to minimize the
 physical resources, handling
increasingly large numbers of quantum system jointly. Regarding this
last point, a relevant issue is the exponentially increasing dimension
of the Choi operator of the
 quantum process versus the number of
systems involved, and methods for safely neglecting irrelevant
parameters in multiple qubit noise model reconstruction have been
introduced \cite{emerson} based on
 assumptions of qubit noise
independence and Markovianity. In Ref. \cite{kosut}, methods to
tackle
 the case of sparse Choi matrices are shown, expressing the
minimum $\ell_1$-norm distance criterion
 in terms of a standard
convex optimization problem. On the problem of optimizing
data-processing, on
 the other hand, upper bounds on minimal
Hilbert-Schmidt distance between the estimated and the
 actual
Choi-Jamio\l kowski state has been derived\cite{scott} exploiting
spherical $2$-designs.  It can
 be shown that minimizing such a
distance is equivalent to the minimizing the statistical error in
 the
estimation of any ensemble average evolved by the channel. On the
other hand, a systematic way
 of posing the problem of optimizing the
data-processing is to fix a cost-function (depending on the
 purpose
of the tomographic reconstruction), and minimize the average
cost---the canonical procedure
 in quantum estimation theory
\cite{helstrom}.  
 
 The optimal data-processing for any measurement
(in finite-dimensions) for estimating the
 expectation of any
observable with minimum error was derived in Ref. \cite{darper}.  On
the other
 hand, in regards of the optimal setups, an approach based
on the theory of quantum combs and quantum
 testers \cite{qca,mem}
have been introduced, that allowed to determine the optimal schemes
(minimizing the statistical error in estimating expectation values)
for all of the three kinds of
 tomography: state, process, and
measurement \cite{opttomo} (quantum combs and quantum testers
generalize the notion of channels and POVM's).  The optimal setups use
up to three ancillas, and
 need only a single input state (with
bipartite entanglement only) and the measurement of a Bell
 basis,
with a variable local unitary shifts of the ancillas. Exploiting the
same approach incomplete
 process tomography has been addressed in
Ref.  \cite{zim} using ``process entropy'', the analogous
 of the
max-entropy method \cite{buz} for process tomography.

\section{methodology}\label{methodology}

In the following we will treat linear operators $X$ from $\sH_0$ to $\sH_1$ as elements of a vector
space, and the following formula is very useful
\begin{equation}
  |X\kk:=\sum_{m=0}^{d_1-1}\sum_{n=0}^{d_0-1}X_{mn}|m\>_1|n\>_0.
  \label{dket}
\end{equation}
In Eq. (\ref{dket}) $d_i$ denotes the dimension of the Hilbert space $\sH_i$, $\{|n\>_i\}$ 
are orthonormal bases for for $\sH_i$ $i=1,2$, and $X_{mn}$ are the
matrix elements of $X$ on the same orthonormal basis.

A general mathematical framework for quantum tomography was introduced in Refs.
\cite{gentomo_ortho,gentomo_quorum}, based on spanning sets of observables called {\em quorums}. In
we will review the more general approach based on {\em informationally complete} POVMs \cite{udet,
  infoc}. A POVM is a set of positive operators $\{P_l\}$ that add up to the identity. The method is
based on operator expansions, and we will show how expanding operators on a POVM can be used to
reconstruct their expectation values on the state of the measured system. The aim of a tomographic
reconstruction is to obtain the ensemble expectation of an operator $X$ by averaging some function
$f_l[X]$ depending on the outcome of a suitable POVM $\{P_l\}$. We require the procedure to be
unbiased, namely the reconstruction must be as follows
\begin{equation}
  \<X\>_\rho=\sum_l f_l[X]p(l|\rho),\quad p(l|\rho):=\Tr[\rho P_l].
  \label{reco}
\end{equation}
Whatever notion of convergence one uses, the requirement for unbiasedeness implies---by the
polarization identity---that the following expansion for the operator $X$ holds
\begin{equation}
  X=\sum_lf_l[X] P_l,
\label{expa}
\end{equation}
where the sum can be replaced by an integral in the case of continuous outcome set (the expansion
clearly is defined for weakly convergent sum, meaning that Eq.~\eqref{reco} holds for all states
$\rho$). The general reconstruction method consists in finding expansion coefficients $f_l[X]$, and
then averaging them over the outcomes $l$. In this way one can define the expansion for general
bounded operators $X$.  Further extensions of the definition in Eq.~\eqref{expa} to unbounded
operators can be obtained requiring the convergence of Eq.~\eqref{reco} for states $\rho$ in a dense
set $\mathcal S$ (e.~g.  finite energy states). A particularly simple case is that of operators on
finite dimensional Hilbert spaces, or for Hilbert-Schmidt operators in infinite dimensional spaces,
since in these cases the space of operators is a Hilbert space itself, equipped with the
Hilbert-Schmidt product $\bb A|B\kk:=\Tr[A^\dag B]$, and convergence of Eq.~\eqref{expa} can be
defined in the Hilbert-Schmidt norm $\n{X}:=\sqrt{\bb X|X\kk}$.\par

Clearly, the use of the formula in Eq.~\eqref{reco} for estimation of
$\<X\>_\rho$ (for all $\rho\in\mathcal S$ and for all $X$ such that
$\Tr[\rho X]$ is defined on $\mathcal S$) is possible iff $\{P_l\}$ is
a complete set in the space of linear operators. Such a POVM is called
informationally complete \cite{bush}. For the sake of simplicity, in
the following we will restrict attention to the case of
Hilbert-Schmidt operators $X$.  Eq.~\eqref{expa} defines a linear map
$\Lambda$ from the vector space of coefficients ${\vect f}:=(f_l)$ to
linear operators as follows
\begin{equation}
  \Lambda{\vect f}=\sum_lf_l P_l,
  \label{lambd}
\end{equation}
whose domain contains all the vectors $\vect f$ such that the sum in
Eq.~\eqref{lambd} converges (either in Hilbert-Schmidt norm or
weakly). As we mentioned before, a reconstruction strategy requires
a choice of coefficients ${\vect f}[X]$ for any operator $X$, such that
$\Lambda{\vect f}[X]=X$. In algebraic terms, the choice corresponds to
a generalized inverse $\Gamma$ of $\Lambda$ defined by
$\Lambda\Gamma\Lambda=\Lambda$, so that ${\vect f}[X]=\Gamma(X)$
\cite{bapat}. When the set $\{P_l\}$ is not linearly independent the
inverse $\Gamma$ is not unique, and this implies that one can choose
the coefficients ${\vect f}[X]$ according to some optimality
criterion, as we will explain in Sec. \ref{s:opt}. Notice that by
linearity, any inverse $\Gamma$ provides a dual spanning set $\{Q_l\}$
whose matrix elements are $(Q_l)^*_{mn}:=f_l[E_{mn}]$, with
$E_{mn}:=|m\>\<n|$, namely $f_l[X]=\Tr[Q^\dag_l X]$.\par

As we will see in the next sections, for finite dimensional systems
the theory of generalized inverses is sufficient for classifying all
possible expansions and consequently deriving the optimal coefficients
${\vect f}[X]$ for a fixed POVM $\{P_l\}$, \cite{darper,magnani}. On
the other hand, the full classification of inverses $\Gamma$ and
consequent optimization is a still unsolved problem for infinite
dimensional systems, for which alternative approaches are useful
\cite{renormtomo}.

\subsection{Frames}

In this subsection we will review the relevant results in the theory of {\em frames} on Hilbert
spaces, which is useful for dealing with POVMs on infinite dimensional systems \cite{renormtomo}
where a classification of all inverses $\Gamma$ is still lacking. The method for evaluating possible
inverses provided in Refs.  \cite{gentomo_ortho,gentomo_quorum} is an orthogonalization
algorithm---similar to the customary Gram-Schmidt method---based on the assumption that the POVM is
a {\em frame} \cite{duffschaeff} in the Hilbert space of Hilbert-Schmidt operators, namely that the
two following inequalities hold
\begin{equation}
  a\n{X}^2\leq\sum_l|\bb P_l|X\kk|^2\leq b\n{X}^2.
\end{equation}
Equivalently, $\{P_l\}$ is a frame iff its {\em frame operator}
\begin{equation}
  F:=\sum_l|P_l\kk\bb P_l|,
  \label{framop}
\end{equation}
is bounded and invertible with bounded inverse. The theory of frames provides a (partial)
classification of inverses $\Gamma$ in terms of {\em dual frames} for $\{P_l\}$, namely those frames
$Q_l$ such that the following identity holds in the vector space of operators
\begin{equation}
  \sum_l|P_l\kk\bb Q_l|=I.
  \label{dual}
\end{equation}
While the orthogonalization method is effective in providing adequate coefficients ${\vect f}[X]$
for the purpose of evaluating the expectation value of operators $X$, it maybe inefficient in
minimizing the statistical errors, since the orthogonalization would be equivalent to discard
experimental data. On the other hand, using the method of alternate duals of a frame allows one to
use all experimental data in the most efficient way, according to any chosen criterion, such as
minimize the statistical error. We will now show how the method works,

The {\em canonical dual} frame is defined as
\begin{equation}
  |D_l\kk:=F^{-1}|P_l\kk,
  \label{candu}
\end{equation}
and it trivially satisfy Eq.~\eqref{dual}. All alternate dual frames of a fixed frame $\{P_l\}$ are
classified in Ref. \cite{li}, and they are given by the following expression
\begin{equation}
  Q_l=D_l+Y_l-\sum_j \bb D_l|P_j\kk Y_j,
  \label{aldu}
\end{equation}
where $Y_l$ is arbitrary, provided that the sum $\sum_j \bb D_l|P_j\kk Y_j$ converges. It is clear
from the definition in Eq.~\eqref{dual} that any dual frame $\{Q_l\}$ corresponds to an inverse
$\Gamma$, via the identification $\Gamma(X)={\vect f}[X]$, with the coefficients given by
\begin{equation}
  f_l[X]:=\bb Q_l|X\kk. 
\end{equation}
For finite dimensions also the converse is true, namely any inverse $\Gamma$ provides a dual set
$\{Q_l\}$ which is a frame. However, in the infinite dimensional case it is not guaranteed that all
the dual sets corresponding to inverses $\Gamma$ are frames themselves.

The results in this subsection can be generalized to frames for
bounded operators (for the theory of frames for Banach spaces, see
Ref. \cite{chl}) by weakening the definition of convergence of the
sums in Eqs.~\eqref{framop}, \eqref{dual}, and \eqref{aldu}.

\section{What you need to measure for tomography}

As we mentioned in the previous section, the use of a detector whose statistics is described by an
informationally complete POVM $\{P_l\}$ allows the reconstruction of any expectation value
(including those of external products $|m\>\<n|$, namely matrix elements in a fixed representation).
In the assumption that every repetition of the experiment is independent, it is indeed sufficient to
find a set of coefficients ${\vect f}[X]$, and to average it by the experimental frequencies
$\nu_l:=n_l/N$ ($n_l$ is the number of outcomes $l$ occurred, and $N$ is the total number of
repetitions). The estimated expectation is then
\begin{equation}
  \overline X:=\sum_l\nu_lf_l[X]\simeq\<X\>_\rho,
  \label{unbias}
\end{equation}
where the symbol $\simeq$ means that by the law of large numbers l.h.s.
converges in probability to r.h.s.

\subsection{Informationally complete measurements}

Informationally complete measurements play a relevant role in foundations of quantum mechanics,
constituting a kind of standard reference measurement with respect to which all quantum quantities
are defined. They have been used as a tool to assess general foundational issues, such as in the
proof of the quantum version of the de Finetti theorem \cite{defi}. One of the most popular examples
of informationally complete measurement is the coherent-state POVM for harmonic oscillators, which
is used in particular in quantum optics. Its probability distribution is the so-called Q-function
(or Husimi function). Other example are the quorums of observables, such as the set of quadratures
of the harmonic oscillator, which was the first kind of informationally complete measurement
considered for quantum tomography \cite{vrisk}. The use of the notion of informational completeness
has also lead to advancements on other relevant conceptual issues, such as the problem of joint
measurements of non-commuting observables \cite{joint}.

\subsection{Quorums}

A quorum of observables $\{X_\xi\}_{\xi \in \mathfrak{X}}$ is a set of
independent observables ($[X_\xi,X_{\xi'}]=0$ only if $\xi=\xi'$), with
spectral resolution $E_\xi(\d x)$ and spectrum ${\mathcal X}_\xi$, such
that the statistics of their outcomes $x\in {\mathcal X}_\xi$ allows
one to reconstruct average values of an arbitrary operator $X$ as
follows
\begin{equation}
  \<X\>_\rho=\int_{\mathfrak X}\mu(\d \xi)\<f_\xi(X_\xi,X)\>_\rho,
  \label{recoquor}
\end{equation}
where $\mu(\d \xi)$ is a probability measure on $\mathfrak X$ and
$f_\xi(x,X)$ is a complex function of $x\in\mathbb{\mathcal X}_\xi$
called {\em tomographic estimator}, enjoying the following properties
\begin{itemize}
\item{In order to have bounded variance in the estimation,
    $f_\xi(x,X)$ is square summable with respect to the measure
    $\mu(\d\xi)\<E_\xi(\d x)\>_\rho$ for all $\rho$ in the set
    $\mathcal S$ of interest, namely
    \begin{equation}
      \int_{\mathfrak X} \mu(\d\xi) \int_{{\mathcal X}_\xi} \<E_\xi(\d x)\>_\rho |f_\xi(x,X)|^2 <\infty,
    \end{equation}
    for $X$ such that $|\Tr[\rho X]|<\infty$ and for all $\rho\in \mathcal S$.}
\item{For a fixed $x$, $f_\xi(x,X)$ is linear in $X$, namely
    \begin{equation}
      \begin{split}
        &f_\xi(x,aX+bY)=af_\xi(x,X)+bf_\xi(x,Y),\\
        &f_\xi(x,X^\dag)=f_\xi(x,X)^*.
      \end{split}
    \end{equation}
  }
\end{itemize}
The problem of tomography is to find all possible correspondences
$X\leftrightarrow f_\xi(x,X)$, namely all possible estimators. Usually
quorums are obtained from observable spanning sets
$\{F_\omega\}_{\omega\in\Omega}$, satisfying
\begin{equation}
  X=\int_\Omega\d\omega\, c_\omega[X]F_\omega,
  \label{spas}
\end{equation}
where the measure $\d\omega$ may be unnormalizable. However, this
feature is usually due to redundancy of the set $\Omega$, which may be
partitioned into sets ${\mathfrak K}_\xi$ of observables such that for
all $F_\omega\in{\mathfrak K}_\xi$, one has $[F_\omega,X_\xi]=0$ for a
fixed observable $X_\xi$. The set ${\mathfrak K}_\xi$ then corresponds
to the observable $X_\xi$ in the quorum so that for
$F_\omega\in{\mathfrak K}_\xi$ we can write
$F_\omega:=F_{(\xi,\kappa)}=f_\kappa(X_\xi)$.  Under standard
hypotheses $\d\omega$ can be decomposed as
$\mu(\d\xi)\nu_\xi(\d\kappa)$, where $\nu_\xi(\d\kappa)$ is the
measure on ${\mathfrak K}_\xi$ induced by $\d\omega$, and
Eq.~\eqref{spas} can be rewritten as
\begin{equation}
  X=\int_{\mathfrak X}\mu(\d\xi)\int_{{\mathfrak K}_\xi}\nu(\d \kappa)c_{(\xi,\kappa)}[X]
  f_\kappa(X_\xi). 
\end{equation}
The last expression has the form of Eq.~\eqref{recoquor} with the
choice of tomographic estimators provided by
\begin{equation}
  f_\xi(x,X):=\int_{{\mathfrak
      K}_\xi}\nu(\d\kappa)c_{(\xi,\kappa)}[X]f_\kappa(x).
  \label{estim}
\end{equation}
Notice that in the case of a quorum the possibility of optimizing the
estimator depends on non uniqueness of the estimator, which is
equivalent to the existence of {\em null functions}, namely functions
$n_\xi(x)$ such that
\begin{equation}
  \int_{\mathfrak X}\mu(\d\xi)\int_{{\mathcal X}_\xi}E_\xi(\d x)n_\xi(x)=0.
\end{equation}

\subsection{Group tomography} 

In this subsection we will review the approach to quantum tomography based on group representations,
that was introduced in Ref.  \cite{tomo_group}, and then exploited in Refs.
\cite{tomo_genova,spintomo}. The method exploits the following group theoretical identity, holding
for unitary irreducible representations $U(g)$ of a unimodular group $G$
\begin{equation}
  \int_G\d g\,U(g)X U(g^\dag)=\Tr[X]I,
  \label{groupav}
\end{equation}
where $\d g$ is the invariant Haar measure of $G$ normalized to 1 [we recall that a group is
unimodular when the left-invariant measure is equal to the right-invariant one]. In the following we
will consider compact Lie groups (such as the rotation group or the group of unitary
transformations), which are necessarily unimodular. However the identity can be extended to square
summable representations of non compact unimodular groups \cite{gromopa}, allowing for extension of
group tomography to the noncompact groups $SU(1,1)$ \cite{tomosu1,su11}, along with the Euclidean
group on the complex plane (which is the case of homodyne tomography). We will exploit the following
identities coming from the correspondence of Eq.~\eqref{dket}
\begin{equation}
  A\otimes B|C\kk=|ACB^T\kk,\quad \Tr_1[|A\kk\bb B|]=A^T B^*,
  \label{dketalg}
\end{equation}
where $X^T$ and $X^*$ denote the transpose and the complex conjugate
of $X$, respectively, on the bases of Eq.~\eqref{dket}. Using
Eqs.~\eqref{groupav} and \eqref{dketalg} one obtains $\int_G\d
g\,|U(g)\kk\bb U(g)|=I$, which implies the following reconstruction
formula
\begin{equation}
  X=\int_G\d g\,\Tr[U^\dag(g)X]U(g).
\end{equation}
In the hypothesis that the group manifold is connected, the
exponential map $e^{i\psi {\vect n}\cdot{\vect T}}$ covers the whole
group, $T_i$ denoting the generators Lie algebra representation and
$\vect n$ being a normalized real vector. The integral can then be
rewritten as follows
\begin{equation}
  X=\int_\Psi\mu(\d\psi)\int_{S^n}\nu(\d{\vect n})\Tr[e^{-i\psi {\vect n}\cdot{\vect T}}X]e^{i\psi {\vect n}\cdot{\vect T}}.
\end{equation}
By exchanging the two integrals over $\psi$ and $\vect n$, the integral over $\psi$ is evaluated
analytically, whereas the integral over $\vect n$ is sampled experimentally. The practical problem
is then to measure ${\vect n}\cdot{\vect T}$. A way is to start from a finite maximal set of
commuting observables, say $\{T_\nu\}$ (these make the so-called Cartan abelian subalgebra of the
Lie algebra), and achieve the observables of the quorum by evolving $T_\nu$ with the group $G$ of
physical transformations in the Heisenberg picture, e.g. by preceding the $T_\nu$-detectors with an
apparatus that performs the transformations of $G$. For example, for the group $SU(2)$ the
generators are the angular momentum components $J_i$, and a quorum is provided by the set of all
angular momentum operators ${\vect J}\cdot{\vect n}$ on the sphere ${\vect n}\in S^2$
\cite{spintomo}, that can be obtained measuring $J_z$ after a rotation of the state.\par

The use of group representations provides also a tool for constructing
{\em covariant} informationally complete POVMs. A covariant POVM with
respect to the representation $U(g)$ of the group $G$ is a POVM with
the following form
\begin{equation}
  P(\d g)=\d gU(g)\xi U(g)^\dag,
\end{equation}
where $\xi \geq 0$ is called {\em seed} and must be such that
$\int_GP(\d g)=I$. The informational completeness can be required
through the invertibility condition for the frame operator in
Eq.~\eqref{framop}, which rewrites
\begin{equation}
  F=\int_G\d g\, U(g)\otimes U(g)^*|\xi\kk\bb\xi| U(g)^\dag\otimes U(g)^T.
\end{equation}
A general classification of covariant informationally complete
measurements has been given in Ref. \cite{infoc}.

\section{Methods of data processing}

Given a detector corresponding to an informationally complete POVM, one can use either the theory of
generalized inverses or the theory of frames to find a suitable data processing to reconstruct all
the parameters of a quantum state. However, the processing is usually not unique, and this feature
leaves room for optimization. One can indeed choose a figure of merit and look for the processing
that optimizes it for a fixed POVM. This step is mandatory for a fair comparison between two POVMs,
and a comparison without optimization generally leads to a wrong choice of POVM.  Before reviewing
recent results on optimization of processing and POVMs, in Sec. \ref{s:opt}, we summarize the main
approaches to data-processing, along with the corresponding figures of merit.

\subsection{The unbiased averaging method: tomography as indirect
  estimation}\label{s:unbi}

Quantum tomography can be regarded as a special case of indirect estimation \cite{joint}, in which
the informationally complete detector allows one to indirectly estimate without bias any expectation
value.  From this point of view, a very natural figure of merit in judging a data processing
strategy is the statistical error in the reconstruction of expectations. The statistical error
occurring when the processing in Eq.~\eqref{expa} is used has the following expression
\begin{equation}
  \Delta(X)_{\rho,{\boldsymbol \nu}}^2:=\sum_l|f_l[X]\nu_l-\<X\>_\rho|^2,
\end{equation}
where the frequencies $\nu_l$ have a multinomial distribution
$p_N({\boldsymbol\nu}|\rho):=\frac{N!}{\prod_l n_l!}\prod_{l}\Tr[\rho
P_l]^{N\nu_l}$. Notice that this reconstruction is unbiased for any
$N$, since averaging the reconstructed expectation in
Eq.~\eqref{unbias} over all possible experimental outcomes provides
exactly $\<X\>_\rho$. On the other hand, averaging the statistical
error over all possible experimental outcomes provides the following
expression
\begin{equation}
  \Delta(X)^2_\rho:=\frac{\sum_l|f_l[X]|^2p(l|\rho)-|\<X\>_\rho|^2}{N}.
\end{equation}
Finally, this quantity depends on the state $\rho$, and in order to
remove this dependence we consider a Bayesian setting in which the
measured state is assumed to be distributed according to a prior
probability $p(\rho)$. Averaging the error over the prior distribution
finally provides
\begin{equation}
  \delta(X)_{\mathcal E}:=\sum_l|f_l[X]|^2p(l|\rho_{\mathcal E})-\overline{|\<X\>|^2}_{\mathcal E},
  \label{figmer}
\end{equation}
where $\rho_{\mathcal E}:=\int\d \rho\, p(\rho)\rho$, and
$\overline{f(\rho)}_{\mathcal E}:=\int\d \rho\, p(\rho)f(\rho)$. In
Refs. \cite{firstopt,infoc}, the expression in Eq.~\eqref{figmer} was
considered as a figure of merit for judging the quality of the
reconstruction provided by the processing coefficients ${\vect f}[X]$
with a fixed POVM $\{P_l\}$. In Sect. \ref{s:opt} we will show how the
optimal processing \cite{darper} can be derived within this
framework.

\subsection{The maximum likelihood method}

The unbiased averaging method can generally lead to expectations that are unphysical, e.g. violating
the positivity of the density operator. This fact had led some authors to adopt data processing
algorithms based on the maximum likelihood criterion, that allows one to constrain the estimated
state to be physical\cite{hrad,banaszek}. However, it actually does not make much difference if the
deviation from the true value results in a physical or unphysical state: is it better to guess a
physical state that is far from the true one, or to guess an unphysical one that is close to the
true one? Indeed, as we have already discussed in Sect. \ref{s:excursus}, the maximum likelihood is
generally biased, and the physical constraint may result e. g. in the state to be pure when instead
the true state is mixed. A Bayesian variation of the maximum-likelihood method was proposed in Ref.
\cite{b-k}, in order to avoid such feature.

A comprehensive maximum-likelihood approach has been given in Ref.  \cite{maxlik}. The likelihood is
a functional $L[\rho]$ over the set of states that evaluates the probability that the state $\rho$
produces the experimental outcomes summarized by the frequencies $\boldsymbol\nu$, and has the
following expression
\begin{equation}
  L[\rho]:=\prod_lp(l|\rho)^{\nu_l}=\left(\prod_lp(l|\rho)^{n_l}\right)^\frac1N.
\end{equation}
It is convenient to define the following functional, which is just the
logarithm of $L[\rho]$
\begin{equation}
  \mathcal L[\rho]:=\frac1N\sum_l n_l \log p(l|\rho),
  \label{loglik}
\end{equation}
whose maximization is equivalent to the maximization of $L[\rho]$. The positivity constraint on
$\rho$ is achieved by substituting it with $T^\dag T$ in Eq.~\eqref{loglik}, thus defining a
functional ${\mathcal L}'[T]$, and introducing a Lagrange multiplier $\mathcal N$ to account for the
condition $\Tr[T^\dag T]=1$.  Eq.~\eqref{loglik} provides a natural interpretation of the maximum
likelihood criterion in terms of the Kullback-Leibler divergence $D({\boldsymbol \nu}|\!|{\vect
  p})$, where $p_l:=p(l|\rho)$. Indeed, the Kullback-Leibler distance of the probability
distribution $\vect p$ from experimental frequencies $\boldsymbol \nu$ has the following expression
\begin{equation}
  D({\boldsymbol\nu}|\!|{\vect p})=\sum_l\nu_l\log\frac{\nu_l}{p_l},
\end{equation}
and since $S({\boldsymbol\nu}):=-\sum_l\nu_l\log\nu_l$ is fixed, the
minimization of the distance is equivalent to the maximization of
\begin{equation}
  \sum_l\nu_l\log p_l=\frac1N\sum_ln_l\log p_l\equiv{\mathcal L}[\rho].
\end{equation}
The maximization over $\rho$ with the positivity and normalization constraints can thus be
interpreted as the choice of a physical state $\rho$ such that its probability distribution has the
minimum Kullback-Leibler distance from the experimental frequencies.\par

The statistical motivation for the maximum likelihood estimator resides in the following argument.
Given a family of probability distributions $p(x;{\boldsymbol \theta})$ in $x$, depending on a
multidimensional parameter $\boldsymbol \theta$, the Fisher information matrix can be defined as
follows
\begin{equation}
  F({\boldsymbol \theta})_{mn}:=\left\<\frac{\partial p(x;{\boldsymbol\theta})}{\partial \theta_m} \frac{\partial p(x;{\boldsymbol\theta})}{\partial \theta_n} \right\>_{x}.
\end{equation}
Upon defining the covariance matrix for an estimator
$\hat{\boldsymbol\theta}$ as follows
\begin{equation}
  \Sigma_{mn}:=\<(\hat \theta_m-\theta_m)(\hat \theta_n-\theta_n)\>_x,
\end{equation}
one has the Cram\'er-Rao bound
\begin{equation}
  \Sigma \geq \frac1NF({\boldsymbol \theta})^{-1},
\end{equation}
which is independent of the estimator $\hat\theta$. It can be proved that when the bound is tight
the maximum-likelihood estimator saturates asymptotically for large $N$. \par

The maximization of the functional ${\mathcal L}[\rho]$ is a nonlinear convex programming problem,
and can be solved numerically. Convergence is assured by convexity and differentiability of the
functional to be maximized over the convex set of states. However, the derivatives of ${\mathcal
  L}[\rho]$ with respect to some of the parameters defining $\rho$ can be very small, so that very
different values of the parameters will give almost the same likelihood, thus making it hard to
judge whether the point reached at a given iteration step is a good approximation of the point
corresponding to the maximum: in such case the problem becomes numerically ill conditioned, with an
extremely low convergence rate.

\subsection{Unbiasing known noise}

In this subsection we will show how the unbiased averaging method explained in Subsect. \ref{s:unbi}
can be applied also in the presence of a known noise disturbing the measurement, provided that the
quantum channel describing the noise is invertible \cite{tomo_group}. The unbiasing method is the
following. Suppose that the noisy channel $\map N$ (in the Heisenberg picture) affects the system
before it is measured by the detector corresponding to the POVM $\{P_l\}$. Then the measured POVM is
actually $\{\map N(P_l)\}$, that for invertible $\map N$ is still informationally complete. The
reconstruction formula Eq.~\eqref{expa} then becomes
\begin{align}
  X&=\map N \map N^{-1}(X)=\map N\left(\sum_lf_l[\map N^{-1}(X)]P_l \right)\nonumber\\
  &=\sum_lf_l[\map N^{-1}(X)]\map N(P_l).
\end{align}
Using the statistics from the measurement of $\{\map N(P_l)\}$ it is then possible to unbias the
noise $\map N$ by averaging the functions ${\vect f}[\map N^{-1}(X)]$. In all known cases, the
coefficients $f_l[Z]$ are obtained as $f_l[Z]=\Tr[Q^\dag_l Z]$ for a dual frame $\{Q_l\}$, and
consequently the coefficients for unbiasing are $\Tr[Q^\dag_l \map N^{-1}(Z)]=\Tr[{\map
  N_*}^{-1}(Q^\dag_l)Z]$, where $\map N_*$ denotes the Schr\"odinger picture of the channel $\map
N$.  As we will see in the following, usually the procedure for unbiasing the noise increases the
statistical error. For examples of noise-unbiasing see Refs.  \cite{Gausstomo,tomo_lecture}.

\section{The quantum systems}


\subsection{Qubits}

The case of a two-dimensional quantum system (qubit) is the easiest example.  Any operator on a
qubit space can be written as
\begin{equation}
  X=\frac12(\Tr[X]I+\sum_{i=1}^3\Tr[X\sigma_i]\sigma_i),
\end{equation}
where $\sigma_i$ are the Pauli matrices. The reconstruction of the expectation $\<X\>_\rho$ can be
obtained by measuring the observables $\sigma_i$ (namely the POVM collecting their eigenstates,
$1/3|\psi_{i\pm}\>\<\psi_{i\pm}|$) and then averaging the function
\begin{equation}
  f_{i\pm}[X]=\frac12(\pm3\Tr[X\sigma_i]+\Tr[X]).
\end{equation}
Also noise unbiasing is particularly easy in this case. Consider for example a depolarizing channel
$\map D_p$ acting in the Heisenberg picture as
\begin{equation}
  \map D_p(X)=(1-p)X+\frac p2\Tr[X]I,
\end{equation}
with $0\leq p<1$. The unbiased estimator is then
\begin{equation}
  f_{i\pm}[X]=\pm\frac3{2(1-p)}\Tr[X\sigma_i]+\frac12\Tr[X].
\end{equation}
The physical realization of a qubit in quantum optics is the {\em dual rail encoding} involving two
modes (typically two different polarization in the same spatial mode) with the logical states
$|0\>_L$ and $|1\>_L$ corresponding to $|0\>|1\>$ and $|1\>|0\>$, respectively.

\subsection{Continuous variables}

The term {\em continuous variables} in the literature has become a synonym of quantum mechanics of a
radiation mode (harmonic oscillator) with creation and annihilation operators $a$ and $a^\dag$. A
spanning set of observables for linear operators on such system is the displacement representation
$D(\alpha):=e^{\alpha a^\dag-\alpha^* a}$ of the Weyl-Heisenberg group, parametrized by
$\alpha\in\mathbb C$, for which the following identity holds
\begin{equation}
  \int_\mathbb C\frac{\d^2\alpha}\pi|D(\alpha)\kk\bb D(\alpha)|=I.
\end{equation}
Notice that we use of the term observable to designate any normal operator $X$ such that
$[X,X^\dag]=0$, so that its real and imaginary parts $(X^\dag+X)/2$ and $i(X^\dag-X)/2$,
respectively, are simultaneously diagonalizable, and unitary operators like $D(\alpha)$ are indeed
normal. The measure $\d^2\alpha$ on the Complex plane $\mathbb C$ is unnormalizable, and plays the
role of the measure $\d\omega$ of Eq.~\eqref{spas}. However, for $\alpha$ with argument
$\arg\alpha=\phi-\pi/2$ we have $[D(\alpha),X_\phi]=[e^{i|\alpha|X_\phi},X_\phi]=0$, where
$X_\phi:=\frac12(a^\dag e^{i\phi}+ae^{-i\phi})$ are the field quadratures. Thus, we can take
$\{D(\alpha)\}$ as the set $\{F_\omega\}$ of Eq.~\eqref{spas}, and the quadratures $X_\phi$ as the
quorum observables $X_\xi$. The integral $\int_{\mathbb C}\frac{\d^2\alpha}\pi$ can be separated as
$\int_0^{\pi}\frac{\d\phi}\pi\int_{-\infty}^{+\infty}\frac{|k|}4\d k$, and since the integral over
$\d k$ is included in the definition of the estimators $f_\phi(X_\phi,X)$ as in Eq.~\eqref{estim},
the remaining integral is the one on $\d\phi$ which is bounded and can be sampled from a uniform
distribution on $[0,\pi)$. The homodyne technique then consists in measuring the informationally
complete POVM $|x\>\<x|_\phi\d x\frac{\d\phi}\pi$ (where $|x\>_\phi$ are Dirac eigenvectors of the
quadrature $X_\phi$), for suitably sampled values of $\phi$, and then averaging the estimators.  The
final reconstruction formula is the following
\begin{equation}
  \<X\>_\rho=\int_0^{\pi}\frac{\d\phi}\pi \int_{-\infty}^\infty\d x\, f_\phi(x,X)\<|x\>\<x|_\phi\>_\rho,
\end{equation}
with $f_\phi(x,X)=\int_{-\infty}^{+\infty}\frac{|k|}4\d
k\,\Tr[D^\dag(ke^{i\phi})X]e^{i kx}$.


\section{Tomography of devices}

Since the publication of Refs. \cite{nielchu,darlop} most of the efforts in
quantum tomography were directed to the reconstruction of devices,
that consists in using the techniques for state reconstruction to the
problem of characterizing the behavior of a quantum device, like a
channel \cite{tomo_hamilt}, a quantum operation
\cite{tomo-rew-spring-lopresti} or a POVM \cite{calipovm}. In the
following subsections we will review the main issues of these
techniques.

\subsection{Tomography of channels}

A quantum channel describes the most general evolution that a quantum system can undergo. It must
satisfy three main requirements: linearity, complete positivity, and preservation of trace (the
physical motivation of complete positivity is that the transformation must preserve positivity of
states also when applied locally to a bipartite system). Probabilistic transformations---so-called
{\em quantum operations}---enjoy linearity and complete positivity, but generally decrease the
trace.\par

The tomography of channels is strictly related to the possibility of imprinting all the information
about a quantum transformation on a quantum state \cite{darlop}, formally expressed by the
Choi-Jamio\l kowski correspondence between a channel $\map{C}:\Lin(\sH_0) \rightarrow \Lin(\sH_1)$
and a positive operator $R_\map C \in \Lin(\sH_1 \otimes \sH_0)$ defined as
\begin{equation}
  R_\map C := (\map{C} \otimes \map{I}) (|I\kk\bb I|),
\end{equation}
where $\map{I}$ is the identity map and $|I\kk \in \sH_0 \otimes
\sH_0$. The correspondence can be inverted as follows
\begin{equation}\label{choiaction}
  \map{C}(\rho) = \Tr_0[(I \otimes \rho^T)R_{\map{C}}],
\end{equation}
and this implies that determining $R_\map C$ is equivalent to determining $\map C$. While complete
positivity of $\map C$ corresponds to positivity of $R_\map C$, trace preservation corresponds to
the condition $\Tr_1[R_\map C]=I$. The reconstruction of the channel $\map C$ can then obtained
preparing the maximally entangled state $1/d(|I\kk\bb I|)$, applying the channel locally and then
reconstructing the output state $d^{-1} R_\map C$. More generally it can be shown that one can use
any bipartite input state $R$ as an input state, as long as it is connected to the maximally
entangled state $1/d(|I\kk\bb I|)$ by an invertible channel \cite{faith}. Such a state is called
{\em faithful}. This situation is actually forced in the infinite dimensional case, where the vector
$|I\kk$ is not normalizable, and e. g. one can use as a faithful state the twin-beam
$T(\lambda)=(1-|\lambda|^2)|\lambda^{a^\dag a}\kk\bb\lambda^{a^\dag a}|$ \cite{faith,altep}.

\subsection{Tomography of measurements}

The statistics and dynamics of a general quantum measurement are described by a quantum instrument,
that is a set of quantum operations $\map E_i$ such that $\sum_i\map E_i=\map E$ is trace
preserving.  Their Choi operators satisfy $R_\map E=\sum_iR_{\map E_i}$, and the POVM describing the
statistics of the measurement is provided by $P_i:=\Tr_1[R_{\map E_i}]$. Similarly to the case of
quantum channels, one can reconstruct quantum operations, along with the whole instrument
corresponding to a measurement \cite{tomo-rew-spring-lopresti}. The tomography of the POVM can be
obtained also for measurements that destroy the system (such as in photo-detection), exploiting the
following argument introduced in Ref.  \cite{calipovm}. If we consider a faithful state $T$, then
measuring the POVM $\{P_i\}$ on $\sH_1$ we have the following conditional state on $\sH_0$
\begin{equation}
  \rho_i:=\frac{\Tr_1[(P_i\otimes I)T]}{\Tr[(P_i\otimes I)T]}.
\end{equation}
Tomographing $\rho_i$ and collecting the statistics of outcomes $i$,
one can reconstruct $P_i$ by inverting the map $\map
T(P)=\Tr[(P\otimes I)T]$ as follows
\begin{equation}
  P_i=\Tr[(P_i\otimes I)T]\map T^{-1}(\rho_i).
\end{equation}

\section{Optimization}\label{s:opt}

In this section we will show the full optimization of quantum
tomographic setups for finite-dimensional states, channels and
measurements, according to the figure of merit defined in
Eq.~\eqref{figmer}. Optimizing quantum tomography is a complex task,
that can be divided in two main steps.\par

The first optimization stage involves a fixed detector, and only
regards the data processing, namely the choice of the inverse $\Gamma$
used to determine the expansion coefficients ${\vect f}[X]$ for a
fixed $X$.  As we will prove in the following, the $\Gamma$ is
independent of $X$, and only depends on the ensemble $\map E$.\par

The second stage consists in optimizing the average statistical error
on a determined set of observables with respect to the POVM, namely
the detector itself.

\subsection{Optimization of data-processing}

In this section we review the data processing optimization, giving the
full derivation in the case of state tomography.
Optimizing the data processing means choosing the best $\Gamma$
according to the figure of merit. As proposed in section \ref{s:unbi},
a natural figure of merit for the estimation of the expectation $\< X
\>_\rho$ of an observable $X$ is the average statistical error; this
is given by the variance $\delta(X)_{\mathcal E}$ of the random
variable $f_l[X]$ with probability distribution $\Tr[\rho_{\mathcal E}
P_i]$, namely $\delta(X)_{\mathcal E}$ defined in Eq.~\eqref{figmer}
The only term in Eq.~\eqref{figmer} that depends on ${\vect f}[X]$ is
$\sum_l|f_l[X]|^2p(l|\rho_{\mathcal E})$, that can be expressed as a
norm in the space $\sK$ of coefficients
\begin{equation}\label{fig-mereta2}
  \sum_l|f_l[X]|^2p(l|\rho_{\mathcal E})=\n{{\vect f}[X]}^2_\pi,
\end{equation}
where $\n{\bf c}^2_\pi:=\sum_{lm}c^*_l\pi_{lm}c_m$, with
\begin{equation}
  \pi_{lm}=\delta_{lm}p(l|\rho_{\mathcal E}).
\end{equation}
It is now clear that minimizing the statistical error in
Eq.~\eqref{figmer} is equivalent to minimizing the norm $\n{{\vect
    f}[X]}_\pi$. In terms of $\pi$ we define the {\em minimum norm}
generalized inverses $\Gamma$: this a generalized inverse that satisfies \cite{magnani}
\begin{equation}
  \pi \Gamma\Lambda=\Lambda^\dag \Gamma^\dag\pi.
\label{minnorm}
\end{equation}
$\Gamma$ has the property that for all $A \in \Rng(\Lambda)$, ${\bf
  f}[A] = \Gamma (A) $ is a solution of the equation $\Lambda {\bf
  f}[A] = A$ with minimum norm.  Notice that the present definition of
minimum norm generalized inverse requires that the norm is induced by a scalar
product (in our case ${\vect a}\cdot{\vect b}:=\sum_{lm}a_l^*\pi_{lm}b_m$).

It can be shown that the minimum norm $\Gamma$ is unique and does not
depend on $X$; the corresponding optimal dual is given by
\cite{darper}
\begin{equation}
 \Gamma =\Lambda^\ddag-([(I-M)\pi(I-M)]^\ddag\pi M)\Lambda^\ddag,
\label{optdual}
\end{equation}
where $M:=\Lambda^\ddag\Lambda$ and $\Lambda^\ddag$ denotes the
Moore-Penrose generalized inverse of $\Lambda$, satisfying
$\Lambda^\ddag\Lambda\Lambda^\ddag=\Lambda^\ddag$ and
$\Lambda^\ddag\Lambda=(\Lambda^\ddag\Lambda)^\dag$. We would like to
stress that as long as the figure of merit can be expresses as a norm
in $\sK$ induced by a scalar product, the optimal processing
represented by $\Gamma$ does not depend on $X$. The minimum of the
expression Eq. (\ref{fig-mereta2}) can be rewritten in this way
\cite{proceqcm}
\begin{equation}\label{eq:YX}
  \delta(X)_{\mathcal E}=\bb X|Y^{-1}|X\kk-\overline{|\<X\>|^2}_{\mathcal E},
\end{equation}
where we defined
\begin{equation}
  Y=\sum_j\frac{\KetBra{P_j}{P_j}}{\Tr[\rho_{\mathcal{E}}P_j]}.
\end{equation}


\subsection{Optimization of the setup}


\subsubsection{Short Review on Quantum Comb Theory}

In this section we give a brief review of the general theory of
quantum circuits, as developed in \cite{qca,mem,smaps}.

\begin{figure}[t]
  \includegraphics[width=\columnwidth]{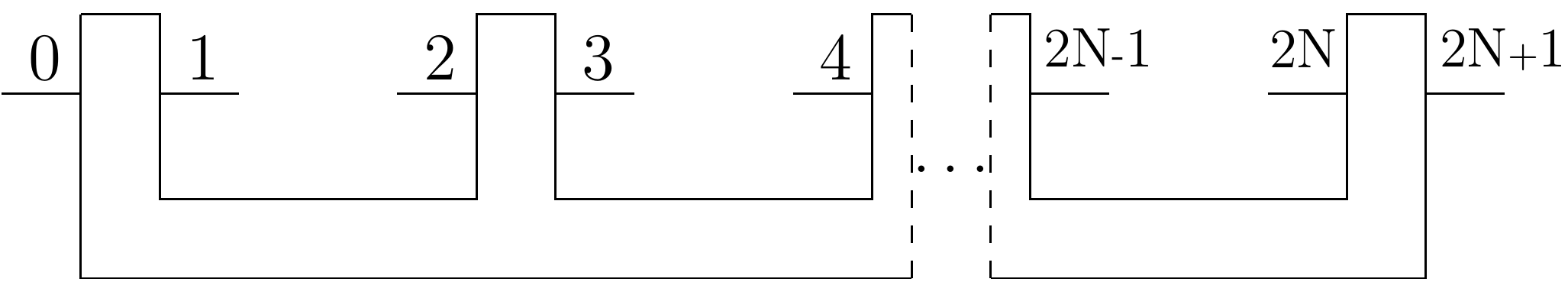}
  \caption{\label{fig:comb} A quantum comb with $N$ slots.
    Information flows from left to right.  The causal structure of the
    comb implies that the input system $m$ cannot influence the output
    system $n$ if $m>n$.}
\end{figure}

\begin{figure}[t]
  \includegraphics[width=\columnwidth]{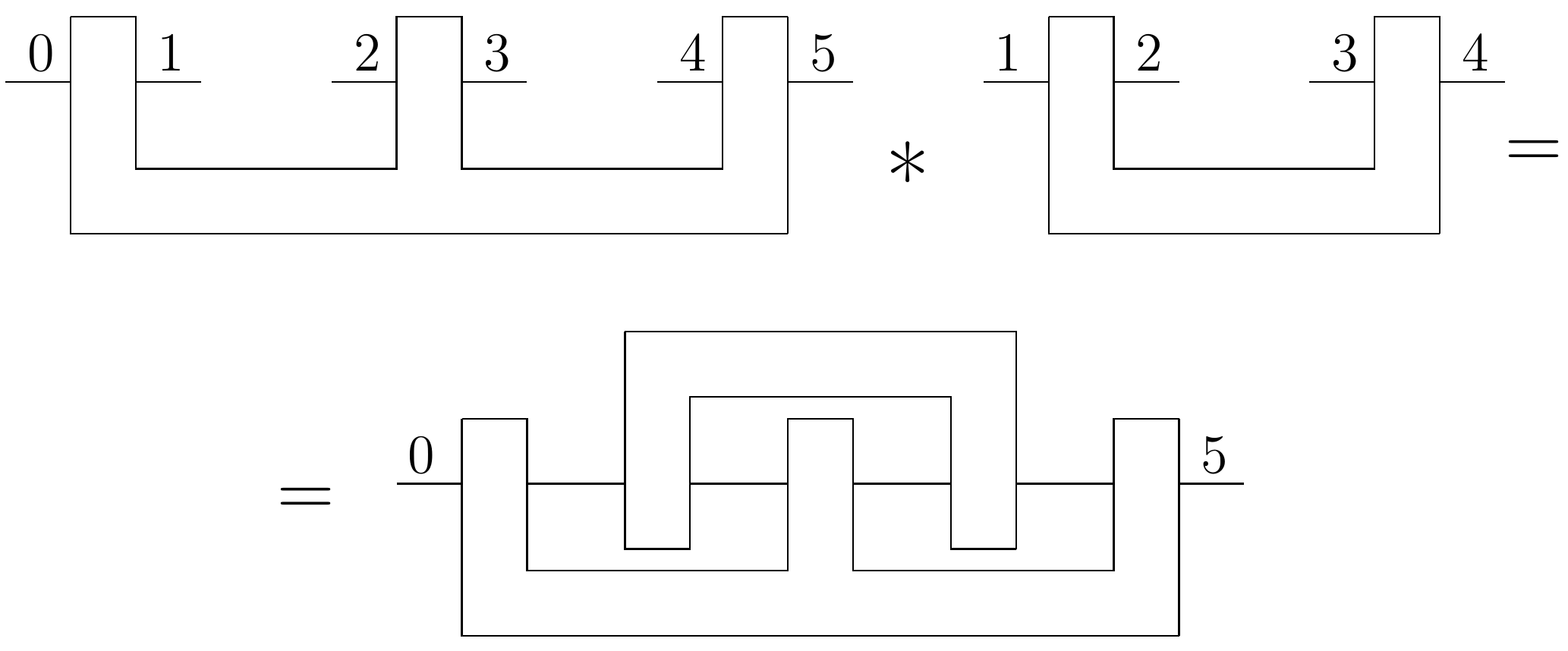}
\caption{\label{fig:comblink} Linking of two combs.  We identify the
  wires with the same label.}
\end{figure}

A quantum comb describes a quantum circuit board, namely a network of
quantum devices with open slots in which variable subcircuits can be
inserted.  A board with $(N-1)$ open slots has $N$ input and output
systems, labeled by even numbers from $0$ to $2N-2$ and by odd
numbers from $1$ to $2N-1$, respectively, as in Fig.  \ref{fig:comb}.
The internal connections of the circuit board determine a {\em causal
  structure}, according to which the input system $m$ cannot influence
the output system $n$ if $m>n$.  Moreover, two circuit boards $\mathscr C_1$
and $\mathscr C_2$ can be connected by linking some outputs of $\mathscr C_1$ with
inputs of $\mathscr C_2$, thus forming a new board $\mathscr C_3 :=\mathscr  C_1 * \mathscr C_2$. We adopt
the convention that wires that are connected are identified by the
same label (see Fig.  \ref{fig:comblink}).

The {\em quantum comb} associated to a circuit board with $N$
input/output systems is a positive operator acting on the Hilbert
spaces $\sH_{out} \otimes \sH_{in}$ where $\sH_{out} :=
\bigotimes_{j=0}^{N-1} \sH_{2j+1}$ and $\sH_{in} := \bigotimes_{j =
  0}^{N-1} \sH_{2j}$, $\sH_n$ being the Hilbert space of the $n$-th
system. For a deterministic circuit board (i.e. a network of quantum
channels) the causal structure is equivalent to the recursive
normalization condition
\begin{equation}\label{recnorm}
\Tr_{2k-1} [ R^{(k)}] = I_{2k-2} \otimes R^{(k-1)} \qquad k=1, \dots, N~
\end{equation} 
where $R^{(N)}=R$, $R^{(0)} =1$, $R^{(k)}
\in\Lin (\bigotimes_{n=0}^{2k-1} \sH_n)$, $\sH_{2n}$ denoting
the Hilbert space of the $n$th input, and $\sH_{2n+1}$ that of the
$n$th output.  We call a positive operator $R$ satisfying
Eq.~\eqref{recnorm}, a \emph{deterministic} quantum comb. We can also
consider \emph{probabilistic} combs, which are defined as the
Choi-Jamio\l kowski operators of probabilistic circuit boards (i.e. network
of quantum operations). A network containing measuring devices will be
then described by a set of probabilistic combs $\{R_i\}$, where the
index $i$ represents a classical outcome. The normalization of
probabilities implies that the sum over all outcomes $R = \sum_i R_i$
has to be a deterministic quantum comb.

The connection of two circuit boards is represented by the {\em link
  product} of the corresponding combs $R_1$ and $R_2$, which is
defined as
\begin{equation}\label{link-prod-def}
  R_1 * R_2 =\Tr_{\sK}[R_1^{\theta_{\sK}} R_2],
\end{equation}
$\theta_{\sK}$ denoting partial transposition over the Hilbert space
$\sK$ of the connected systems (recall that we identify with the same
labels the Hilbert spaces of connected systems).  Note that Eq.
\eqref{choiaction}, which gives the action of a channel $\mathscr{C}$
on a state $\rho$ in term of the Choi operator $C$, can be rewritten
using the link product as $\mathscr{C}(\rho) = C*\rho$.  Moreover,
when variable circuits with Choi operators $ C_j \in \Lin (\sH_{2j}
\otimes \sH_{\sH_{2j-1}}), ~ j=1, \dots, N-1$ are inserted as inputs
in the slots of the circuit board, one obtains as output the quantum
operation $\mathscr C'$ given by
\begin{equation}
C' =  R * C_1 * C_2 * \dots * C_{N-1}~.
\end{equation}


According to the above equation, quantum combs describe all possible
manipulations of quantum circuits, thus generalizing the notions of
quantum channel and quantum operation to the case of transformations
where the input is not a quantum system, but rather a set of quantum
operations.  An important example of such transformations is that of
\emph{quantum testers}, i.e.  transformations that take circuits as
the input and provide probabilities as the output.  A tester is a set
of probabilistic combs $\{\Pi_i\}$ with one-dimensional spaces $\sH_0$
and $\sH_{2N-1}$, with the sum $\Pi = \sum_i \Pi_i$ being a
deterministic comb satisfying Eq.  (\ref{recnorm}).  When connecting
the tester with another circuit board $R$ we obtain the probabilities
$p_i = \Pi_i * R = \Tr [\Pi_i^T R]$, which, a part from the transpose
(which can be reabsorbed into the definition of the tester), is
nothing but the generalization of the Born rule for quantum networks.
In the particular case of testers with a single slot, the tester is a
set of probabilistic combs $\{ \Pi_i \in \Lin(\sH_{out} \otimes
\sH_{in})\}$, and its normalization becomes
\begin{equation}\label{eq-tester}
\sum_i \Pi_i = I_{out} \otimes \sigma, \quad \sigma \ge 0, \Tr[\sigma] = 1.
\end{equation}
When connecting a channel $\mathscr C$ to the tester, the latter provides the
outcome $i$ with probability
\begin{equation}\label{Borntester}
p_i=\Tr[R_{{\mathcal C}}\Pi_i]~,
\end{equation}
where $R_{{\mathcal C}}$ is the Choi operator of $\mathcal{C}$.

It is easy to see that every tester $\{\Pi_i\} $ can be realized with
the following physical scheme: \emph{i}) prepare the pure state
$|\sqrt{\sigma^T} \kk \in \Lin (\sH_{in}\otimes\sH_{in})$, \emph{ii)}
apply the channel $\mathscr C$ on one side of the entangled state
\emph{iii}) measure the joint POVM $\{ P_i = \Pi^{- \frac 1 2} \Pi_i
\Pi^{- \frac 1 2}\}$, where $\Pi^{- 1}$ is the g-inverse $\Pi$. With this  scheme one has indeed
\begin{equation}
p_i = \Tr [ P_i   (\mathscr C \otimes I) (|\sqrt{\sigma^T} \kk \bb \sqrt{\sigma^T} |) ] = \Tr [\Pi_i R_{\mathscr C} ]~. 
\end{equation}

Tomographing a quantum transformation means using a suitable
tester $\Pi_i$ such that the expectation value of any other possible
measurement can be inferred by the probability distribution $p_i =
\Tr[R_{{\mathcal T}} \Pi_i]$.  In order to achieve this task we have
to require that $\{ \Pi_i \}$ is an operator frame for
$\Lin(\sH_{out} \otimes \sH_{in})$.  This means that we can
expand any operator on $\sH_{out} \otimes \sH_{in}$ as follows
\begin{equation}\label{eq-operator}
  A = \sum_l \bb\Delta_l|A\kk \Pi_l \qquad A \in \mathcal{B}(\sH_{out} \otimes  \sH_{in}),
\end{equation}
where we use the fact that for all generalized inverses $\Gamma$ one
has $f_l[X]=\Tr[\Delta^\dag_l X]$ with $\{\Delta_l\}$ a possible dual
spanning set of $\{ \Pi_l \}$ satisfying the condition $\sum_i
|\Pi_i\kk\bb\Delta_i| = I_{out} \otimes I_{in}$.

Optimizing the tomography of quantum transformations means minimizing
the statistical error in the determination of the expectation of a
generic operator $A$ as in Eq. (\ref{eq-operator}). The optimization
of the dual frame follows exactly the same lines as for state
tomography and gives the same result of Eq.~\eqref{eq:YX}, provided
that i) the POVM $\{ P_i \}$ is replaced by the tester $\{ \Pi_i \}$
ii) the ensemble $\mathcal E$ becomes an ensemble $\mathcal{E} =
\{R_k, p_k \}$ of possible transformations and the average state
$\rho_{\mathcal{E}}$ becomes the average Choi operator
$R_\mathcal{E}$.

\subsubsection{Derivation of the optimized setup}

In this section we address the problem of the optimization of the
tester $\{ \Pi_i \}$. A priori one can be interested in some
observables more than other ones, and this can be specified in terms
of a weighted set of observables $\mathcal{G} = \{X_n , q_n \}$, with
weight $q_n>0$ for the observable $X_n$. The optimal tester depends on
the choice of $\mathcal G$, as we will prove in the following. We can
assume that we already optimized the data-processing, so that the
minimum statistical error averaged over $\mathcal{G}$, leading to
\begin{equation}\label{eq-figmersg}
  \delta_{\mathcal{E},\mathcal{G}} := \sum_n \Bra{X_n}Y^{-1}\Ket{X_n} - \sum_{n} q_n \overline{|\<X_n\>|^2}_{\mathcal E}.
\end{equation}
Notice that only the first addendum of Eq. (\ref{eq-figmersg}) depends
on the tester, so we just have to minimize
\begin{equation}\label{eq:YG}
  \eta_{\mathcal{E},\mathcal{G}} := \Tr[Y^{-1}G],
\end{equation}
where $G=\sum_n q_n \KetBra{X_n}{X_n}$. 



In the following, for the sake of clarity we will consider
$\hbox{dim}(\sH_{1})=\hbox{dim}(\sH_{2})=:d$, and focus on the
``symmetric'' case $G = I$; this happens for example when the set $\{
X_n \}$ is an orthonormal basis, whose elements are equally weighted.
Moreover, we assume that the averaged channel of the ensemble
$\mathcal{E}$ is the maximally depolarizing channel, whose Choi
operator is $R_\mathcal{E}=d^{-1}I\otimes I$. Since $R_\mathcal{E}$ is
invariant under the action of $SU(d) \times SU(d)$ we now show that it
is possible to impose the same covariance also on the tester without
increasing the value of $\eta_{{\mathcal E},{\mathcal G}}$.  Let us define
\begin{align} \label{eq-testcovar}
\Pi_{i,g,h}& := (U_g\otimes V_h) \Pi_i (U_g^\dag \otimes V_h^\dag),\\
\Delta_{i,g,h}& := (U_g\otimes V_h) \Delta_i (U_g^\dag \otimes V_h^\dag).
\end{align}
It is easy to check that $\Delta_{i,g,h}$ is a dual of $\Pi_{i,g,h}$.
In fact, using identity in Eq. (\ref{dketalg}), we have
\begin{align}
  & \sum_i \int\!\d g \d h \, |\Pi_{igh}\kk\bb \Delta_{igh}| = \\
  & \int\! \d g \d h \, W_{gh} \left(\sum_i |\Pi_{i}\kk\bb \Delta_{i}|\right)
  W^\dag_{gh} = d^{-1} I \otimes I
\end{align}
where $\d g$ and $\d h$ denote the Haar measure normalized to unit,
and $W_{gh} := (U_g \otimes V_h) \otimes (U^*_g \otimes V^*_h)$. Then
we observe that the normalization of $\Pi_{i,g,h}$ gives
\begin{equation}
\sum_i \! \int \!\! \d g \d h \; \Pi_{i,g,h} = d^{-1}I \otimes I
\end{equation}
corresponding to $\sigma= d^{-1}I$ in Eq. (\ref{eq-tester}), namely
one can choose $\nu=d^{-1}|I\kk\bb I|$. It is easy to verify that the
figure of merit for the covariant tester is the same as for the non
covariant one, whence, w.l.o.g. we optimize the covariant tester. The
condition that the covariant tester is informationally complete w.r.t.
the subspace of transformations to be tomographed will be verified
after the optimization.

We note that a generic covariant tester is obtained by Eq.
(\ref{eq-testcovar}), with operators $\Pi_i$ becoming seeds of the
covariant POVM, and now being required to satisfy only the
normalization condition
\begin{equation} \label{eq-norm} 
  \sum_i \Tr[\Pi_i]= d
\end{equation}
(analogous of covariant POVM normalization in \cite{infoc,holevo}).
With the covariant tester and the assumptions $G = I$,
$R_{\mathcal{E}}=I$  Eq.~(\ref{eq:YG}) becomes
\begin{equation}\label{e:eta0}
  \eta_{{\mathcal E},{\mathcal G}} = \Tr[\tilde{Y}^{-1}],
\end{equation}
where
\begin{align}
 \tilde{Y}  = \sum_i \int \!\! \d g\d h \; \frac{ d |\Pi_{i,g,h}\kk\bb \Pi_{i,g,h}| }{\Tr[\Pi_{i,g,h}]}=
\int \!\! \d g\d h \; W_{g,h}
 X W_{g,h}^\dag
\end{align}
with $Y = \sum_i d|\Pi_i\kk\bb\Pi_i|/\Tr[\Pi_i]$.
Using Schur's lemma we have
\begin{align}
&\tilde{Y}=P_1 + A P_2 + B P_3 + C P_4,\label{eq-projcov}\\
&\!\!\begin{array}{ll}
P_1= \Omega_{13} \otimes \Omega_{24},
&P_2= \left(I_{13} - \Omega_{13}\right) \otimes \Omega_{24},\\
P_3= \Omega_{13} \otimes \left(I_{24} - \Omega_{24}\right),
&P_4= (I_{13} - \Omega_{13}) \otimes (I_{24} - \Omega_{24}),\nonumber
\end{array}
\end{align}
having posed $\Omega= {|I\kk\bb I|}/{d}$ and
\begin{align}
&A = \frac{1}{d^2-1} \left\{\sum_i\frac{\Tr[(\Tr_2[\Pi_i])^2]}{\Tr[\Pi_i]}-1\right\},\nonumber\\
&B = \frac{1}{d^2-1} \left\{\sum_i\frac{\Tr[(\Tr_1[\Pi_i])^2]}{\Tr[\Pi_i]}-1\right\},\\
&C = \frac{1}{(d^2-1)^2}\left\{\sum_i \frac{d\Tr[\Pi_i^2]}{\Tr[\Pi_i]}-(d^2-1)(A+B)-1\right\}.\nonumber
\end{align}
One has
\begin{equation}\label{eq-trx}
\Tr[\tilde{Y}^{-1}]= 1 + (d^2-1) \left( \frac{1}{A}+\frac{1}{B} + \frac{(d^2-1)}{C}\right).
\end{equation}

Notice that if the ensemble of transformations is contained in a
subspace $\mathcal{V} \subseteq
\mathcal{B}(\sH_{2}\otimes\sH_{2})$ the figure of merit
becomes $\eta= \Tr[\tilde{Y}^\ddagger Q_\mathcal{V}]$. We now
carry on the minimization for three relevant subspaces:
\begin{align}
& \mathcal{Q} = \mathcal{B}(\sH_{2}\otimes\sH_{1}),\qquad
\mathcal{C} = \{ R \in \mathcal{Q}, \; \Tr_{2}[R]=I_{1} \} & \nonumber  \\
& \mathcal{U} = \{ R\in \mathcal{Q} , \; \Tr_{2}[R]=I_{1}, \Tr_{1}[R]=I_{2} \}&
\end{align}
corresponding respectively to quantum operations, general channels and
unital channels. The subspaces $\mathcal{C}$ and $\mathcal{U}$ are
invariant under the action of the group $\{ W_{g,h} \}$ and thus the
respective projectors decompose as
\begin{equation}
Q_\mathcal{C} = P_1 + P_2 + P_4, \qquad Q_\mathcal{U} = P_1 + P_4
\end{equation}


Without loss of generality we can assume the operators $\{ \Pi_i \}$
to be rank one. In fact, suppose that $\Pi_i$ has rank higher than 1.
Then it is possible to decompose it as $\Pi=\sum_{j}\Pi_{i,j}$ with
$\Pi_{i,j}$ rank 1. The statistics of $\Pi_i$ can be completely
achieved by $\Pi_{i,j}$ through a suitable post-processing. For the
purpose of optimization it is then not restrictive to consider rank
one $\Pi_i$, namely $\Pi_i=\alpha_i|\Psi_i\kk\bb\Psi_i|$, with
$\sum_i\alpha_i=d$.  Notice that all multiple seeds of this form lead
to testers satisfying Eq. (\ref{eq-norm}). In the three cases under
examination, the figure of merit is then
\begin{align}
&\eta_\mathcal{Q}=\Tr[\tilde Y^{-1}]=1+(d^2-1)\left(\frac2A+\frac{(d^2-1)^2}{1-2A}\right) & \nonumber \\
&\eta_\mathcal{C}=\Tr[\tilde Y^\ddagger Q_\mathcal{C}] =1+(d^2-1)\left(\frac1A+\frac{(d^2-1)^2}{1-2A}\right) & \nonumber \\
&\eta_\mathcal{U}=\Tr[\tilde Y^\ddagger Q_\mathcal{U}] =1+(d^2-1)\left(\frac{(d^2-1)^2}{1-2A}\right) &
\end{align}
where $0\leq
A=(d^2-1)^{-1}(\sum_i\alpha_i\Tr[(\Psi_i\Psi_i^\dag)^2]-1) \leq\frac1{d+1}<\frac12$.
The minimum can simply be determined by derivation with respect to
$A$, obtaining $A=1/(d^2+1)$ for quantum operations,
$A=1/(\sqrt2(d^2-1)+2)$ for general channels and $A=0$ for unital
channels. The corresponding minimum for the figure of merit is
\begin{align}
&\eta_\mathcal{Q} \geq d^6+d^4-d^2 & \nonumber \\
&\eta_\mathcal{C} \geq d^6+(2\sqrt2-3)d^4+(5-4\sqrt2)d^2+2(\sqrt2-1) & \nonumber \\
&\eta_\mathcal{U} \geq (d^2-1)^3+1.&
\end{align}
The same result for quantum operations and for unital channels has
been obtained in \cite{scott2} in a different framework.

These bounds are simply achieved by a single seed
$\Pi_0=d|\Psi\kk\bb\Psi|$, with
\begin{equation}
 \Tr[(\Psi\Psi^\dag)^2] = \frac{2d}{d^2+1},\quad
 \frac{\sqrt2(d^2-1)+1+d^2}{d(\sqrt2(d^2-1)+2)},\quad \frac1d 
\end{equation}
respectively for quantum operations, general channels and unital channels,
namely with
\begin{equation}\label{eq-psi}
\Psi=[d^{-1}(1-\beta) I+\beta|\psi\>\<\psi|]^{\frac12}
\end{equation}
where $\beta=[(d+1)/(d^2+1)]^{1/2}$ for quantum operations,
$\beta=[(d+1)/(2+\sqrt2(d^2-1))]^{1/2}$ for general channels and
$\beta=0$ for unital channels,
and $|\psi\>$ is any pure state. The informational completeness
is verified if the operator
\begin{equation}
F=\int \!\! \d g\d h \, |\Pi_{0gh}\kk\bb\Pi_{0gh}|
\end{equation}
is invertible, namely (see \cite{infoc}) if, for every $i$,
\begin{equation}
  \bb\Psi|\bb\Psi| P_i |\Psi\kk|\Psi\kk \neq 0,
\end{equation}
which is obviously true for $\Psi$ defined in Eq. (\ref{eq-psi}).

The same procedure can be carried on when the operator $G$ has the
more general form $G=g_1 P_1 + g_2 P_2 + g_3 P_3 + g_4 P_4$, where
$P_i$ are the projectors defined in (\ref{eq-projcov}). In this case
Eq. (\ref{eq-trx}) becomes
\begin{equation}
\Tr[\tilde{Y}^{-1}G]= g_1 + (d^2-1) \left( \frac{g_2}{A}+\frac{g_3}{B} + \frac{(d^2-1)g_4}{C}\right),
\end{equation}
which can be minimized along the same lines previously followed.  $G$
has this form when optimizing measuring procedures of this kind:
\emph{i}) preparing an input state randomly drawn from the set $\{U_g
\rho U^{\dag}_g\} $; \emph{ii}) measuring an observable chosen from
the set $\{U_h A U^{\dag}_h\} $. With the same derivation, but keeping
$\dim(\sH_{1})\neq\dim(\sH_{2})$, one obtains the optimal
tomography for general quantum operations. The special case of
$\text{dim}(\sH_{2})=1$ (one has $P_3=P_4=0$ in Eq.
(\ref{eq-projcov})) corresponds to optimal tomography of states,
whereas case $\text{dim}(\sH_{2})=1$ ($P_2=P_4=0$) gives the optimal
tomography of POVMs.

\subsubsection{Experimental realization schemes}

\begin{figure}[t]
\begin{center}
\includegraphics[width=.7\columnwidth]{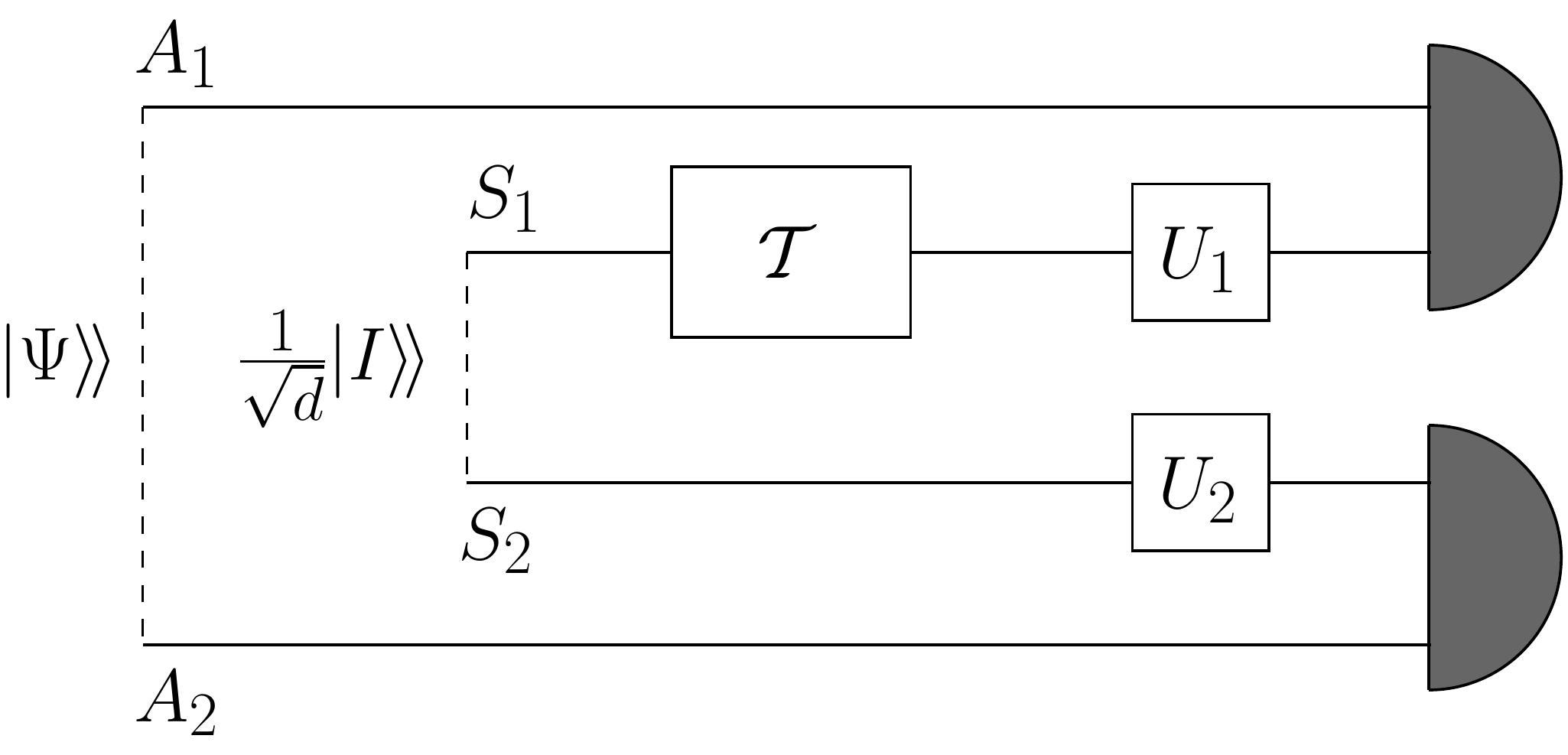}
\end{center}
\caption{ \label{optimalfig} Physical implementation of optimal
  quantum transformation tomography.  The two measurements are Bell's
  measurements preceded by a random unitary. The state $|\Psi\kk$
  depends on the prior ensemble.}
\end{figure}

We now show how the optimal measurement can be experimentally
implemented. Referring to Fig.  \ref{optimalfig}, the bipartite system
carrying the Choi operator of the transformation is indicated with the
labels $S_1$ and $S_2$. We prepare a pair of ancillary systems $A_1$
and $A_2$ in the joint state $|\Psi\kk\bb\Psi|$, then we apply two
random unitary transformations $U_1$ and $U_2$ to $S_1$ and $S_2$,
finally we perform a Bell measurement on the pair $A_1 S_1$ and
another Bell measurement on the pair $A_2 S_2$. This experimental
scheme realizes the continuous measurement by randomizing among a
continuous set of discrete POVM; this is a particular application of a
general result proved in \cite{contmeasure}. The scheme proposed is
feasible using e.~g. the Bell measurements experimentally realized in
\cite{zeilinger}.  We note that choosing $|\Psi\kk$ maximally
entangled (as proposed for example in \cite{mohlid3}) is generally not
optimal, except for the unital case.

The experimental schemes for POVMs/states are obtained by removing the
upper/lower for branch quantum operations, respectively. In the
remaining branch the bipartite detector becomes mono-partite,
performing a von Neumann measurement for the qudit, preceded by a
random unitary in $SU(d)$.  Moreover, for the case of POVM, the state
$|\Psi\kk$ is missing, whereas, for state-tomography, both bipartite
states are missing. The optimal $\eta_{{\mathcal E},{\mathcal G}}$ in
Eq. (\ref{e:eta0}) is given by $\eta=d^3+d^2-d$, in both cases (for
state-tomography compare with Ref. \cite{scott}).

\vfill
\vfill












\begin{thebibliography}{99}
\bibitem{libroparis} M. G. A. Paris and J. \v Reh\'a\v cek (Eds.) {\em
    Quantum State Estimation}, Lect. Notes Phys. {\bf 649} (Springer,
  Berlin-New York, 2004).
\bibitem{review_cerf} G. M. D'Ariano, L. Maccone, and M. F.  Sacchi, {\em Homodyne tomography and
    the reconstruction of quantum states of light,} in {\em Quantum Information with Continuous
    Variables of Atoms and Light}, Ed. by N. Cerf, G. Leuchs, and E. Polzik, (Imperial College
  Press, London, 2007).
\bibitem{revtomo} G. M. D'Ariano, M. G. A. Paris, M. F. Sacchi,
  {\em Quantum~Tomography}, Advances in Imaging and Electron Physics
  {\bf 128} 205-308 (2003)
\bibitem{bush} P. Busch, \emph{Informationally complete sets of physical quantities}, Int. J. Theor.
  Phys. {\bf 30}, 1217 (1991).
\bibitem{udet} G. M. D'Ariano, P. Perinotti, and M. F. Sacchi, {\em Quantum Universal Detectors},
  Europhys.  Lett. {\bf 65} 165 (2004)
\bibitem{infoc} G. M. D'Ariano, P. Perinotti, and M. F. Sacchi, \emph{Informationally complete
    measurements and group representation}, J.  Opt. B: Quantum Semiclass. Opt. {\bf 6}, S487-S491
  (2004).
\bibitem{darper} G. M. D'Ariano and P. Perinotti, {\em Optimal Data Processing for Quantum
    Measurements}, Phys. Rev.  Lett.  {\bf 98}, 020403 (2007).
\bibitem{opttomo} A. Bisio, G. Chiribella, G. M. D'Ariano, S.
  Facchini, and P. Perinotti, {\em Optimal Quantum Tomography of
    States, Measurements, and Transformations}, Phys.  Rev. Lett. {\bf
    102}, 010404 (2009).
\bibitem{qca} G. Chiribella, G. M. D'Ariano, and P. Perinotti, {\em
    Quantum Circuit Architecture}, Phys.  Rev. Lett. {\bf 101}, 060401
  (2008).
\bibitem{fano} U. Fano, {\em Description of States in Quantum
    Mechanics by Density Matrix and Operator Techniques}, Rev. Mod.
  Phys. {\bf 29}, 74 (1957).
\bibitem{pauli} W. Pauli, in {\em Encyclopedia of Physics} {\bf V}
  (Springer, Berlin 1958) p.~17.
\bibitem{bandpark} W. Band and J. L. Park, {\em The empirical
    determination of quantum states}, Found. Phys. {\bf 1}, 133
  (1970).
\bibitem{despag} B. D'Espagnat, {\em Conceptual Foundations of Quantum Mechanics} (W. A. Benjamin,
 Mass. 1976).
\bibitem{buz} V. Bu\v zek, \emph{Quantum tomography from incomplete data via MaxEnt principle}, in Lect. Notes Phys. {\bf 649} 189 (2004).
\bibitem{olipa} S. Olivares and M. G. A. Paris,
 \emph{Quantum estimation
    via the minimum Kullback entropy principle},
 Phys. Rev. A {\bf
    76}, 042120 (2007).
\bibitem{Royer} A. Royer, {\em Measurement of quantum states and the
    Wigner function}, Found. Phys. {\bf 19}, 3 (1989).
\bibitem{reim} D. T. Smithey, M. Beck, M. G. Raymer, and A. Faridani,
  {\em Measurement of the Wigner distribution and the density matrix
    of a light mode using optical homodyne tomography - Application to
    squeezed states and the vacuum}, Phys. Rev. Lett. {\bf 70}, 1244
  (1993).
\bibitem{vrisk} K. Vogel and H. Risken, {\em Determination of
    quasiprobability distributions in terms of probability
    distributions for the rotated quadrature phase}, Phys. Rev. A {\bf
    40}, 2847 (1989).
\bibitem{dar} G. M. D'Ariano, C. Macchiavello, and M. G. A. Paris,
  {\em Detection of the density matrix through optical homodyne
    tomography without filtered back projection}, Phys. Rev. A {\bf
    50}, 4298 (1994).
\bibitem{ulf2} U. Leonhardt, H. Paul and G. M. D'Ariano, {\em
    Tomographic reconstruction of the density matrix via pattern
    functions}, Phys. Rev. A {\bf 52}, 4899 (1995).
\bibitem{bilktomo} G. M. D'Ariano, {\em Measuring Quantum States}, in
  {\em Quantum Optics and Spectroscopy of Solids}, ed. by T.
  Hakio\v{g}lu and A. S. Shumovsky, (Kluwer Academic Publisher,
  Amsterdam, 1997), p. 175-202.
\bibitem{raymer95} M. Munroe, D. Boggavarapu, M.~E. Anderson, and
  M.~G. Raymer, {\em Photon-number statistics from the phase-averaged
    quadrature-field distribution: Theory and ultrafast measurement},
  Phys. Rev. A {\bf 52}, R924 (1995).
\bibitem{exper-onemode2} S. Schiller, G. Breitenbach, S.~ F. Pereira,
  T. M\"{u}ller and J. Mlynek, {\em Quantum statistics of the squeezed
    vacuum by measurement of the density matrix in the number state
    representation}, Phys. Rev. Lett. {\bf 77} 2933 (1996); G.
  Breitenbach, S. Schiller and J. Mlynek, {\em Measurement of the
    quantum states of squeezed light}, Nature {\bf 387}, 471 (1997).
\bibitem{atobeams} U. Janicke and M. Wilkens, {\em Tomography of Atom
    Beams}, J. Mod. Opt. {\bf 42}, 2183 (1995); S. Wallentowitz and W.
  Vogel, {\em Reconstruction of the Quantum Mechanical State of a
    Trapped Ion}, Phys. Rev. Lett. {\bf 75}, 2932 (1995); S. H. Kienle,
  M.  Freiberger, W. P. Schleich, and M. G.  Raymer in {\em
    Experimental Metaphysics: Quantum Mechanical Studies for Abner
    Shimony} ed. by S. Cohen et al. (Kluwer, Lancaster, 1997) p. 121.
\bibitem{diatomolecu} T. J. Dunn, I. A. Walmsley and S. Mukamel, {\em
    Experimental Determination of the Quantum-Mechanical State of a
    Molecular Vibrational Mode Using Fluorescence Tomography}, Phys.
  Rev. Lett. {\bf 74}, 884 (1995).
\bibitem{exper-freeatoms} C. Kurtsiefer, T. Pfau T. and J. Mlynek,
  {\em Measurement of the Wigner function of an ensemble of helium
    atoms}, Nature {\bf 386}, 150 (1997).
\bibitem{leibfried} D. Leibfried, D. M. Meekhof, B. E. King, C.
  Monroe, W. M. Itano and D. J. Wineland, {\em Experimental
    Determination of the Motional Quantum State of a Trapped Atom},
  Phys. Rev. Lett. {\bf 77}, 4281 (1996).
\bibitem{twoself} G. M. D'Ariano, M. Vasilyev, and P. Kumar, {\em
    Self-homodyne tomography of a twin-beam state}, Phys.  Rev. A {\bf
    58}, 636 (1998).
\bibitem{tokyo} G. M. D'Ariano, {\em Homodyning as universal
    detection}, in {\em Quantum Communication, Computing, and
    Measurement}, edited by O. Hirota, A. S. Holevo, and C. M. Caves
  (Plenum Publishing, New York and London, 1997) p.~253.
\bibitem{1LO} G. D'Ariano, P. Kumar, M. Sacchi, {\em Universal
    homodyne tomography with a single local oscillator}, Phys. Rev. A
  {\bf 61}, 13806 (2000).
\bibitem{chicago} G. M. D'Ariano, {\em Latest developements in quantum
    tomography}, in {\em Quantum Communication, Computing, and
    Measurement}, edited by P. Kumar, G. M. D'Ariano, and O. Hirota
  (Kluwer Academic/Plenum Publishers, New York and London, 2000)
  p.~137.
\bibitem{tomo_group} G. M. D'Ariano, {\em Universal quantum
    estimation}, Phys. Lett. A {\bf 268}, 151 (2000).
\bibitem{tomo_genova} G. Cassinelli, G. M. D'Ariano, E. De Vito, A.
  Levrero, {\em Group TheoreticalQuantum Tomography}, J. Math. Phys.
  {\bf 41}, 7940 (2000)
\bibitem{gentomo_ortho} G. M. D'Ariano, L. Maccone, and M. G. A.
  Paris, {\em Orthogonality relations in Quantum Tomography}, Phys.
  Lett. A {\bf 276}, 25 (2000)
\bibitem{gentomo_quorum} G. M. D'Ariano, L. Maccone and M. G. A.
  Paris, {\em Quorum of observables for universal quantum estimation},
  J. Phys. A: Math. Gen. {\bf 34}, 93 (2001).
\bibitem{prugo} E. Prugove\v{c}ki, {\em Information-theoretical
    aspects of quantum measurement},Int. J. Theor. Phys {\bf 16}, 321
  (1977).
\bibitem{hrad} Z. Hradil, {\em Quantum-state estimation}, Phys. Rev. A
  {\bf 55}, R1561 (1997).
\bibitem{banaszek} K. Banaszek, {\em Maximum-likelihood estimation of
    photon-number distribution from homodyne statistics} Phys. Rev. A
  {\bf 57}, 5013 (1998).
\bibitem{maxlik} K. Banaszek, G. M. D'Ariano, M. G. A. Paris, M. F.
  Sacchi, {\em Maximum-likelihood estimation of the density matrix},
  Phys. Rev. A {\bf 61} 010304(R) (2000).
\bibitem{b-k} R. Blume-Kohout, {\em Optimal, reliable estimation of
    quantum states}, quant-ph/0611080.
\bibitem{nielchu} I. L. Chuang and M. A. Nielsen, {\em Prescription
    for experimental determination of the dynamics of a quantum black
    box}, J. Mod. Opt. {\bf 44}, 2455 (1997).
\bibitem{poya} J. F. Poyatos, J. I. Cirac, and P. Zoller, {\em
    Complete Characterization of a Quantum Process: The Two-Bit
    Quantum Gate}, Phys. Rev.  Lett. {\bf 78}, 390 (1997).
\bibitem{tomo_hamilt} G. M. D'Ariano and L. Maccone, {\em Measuring
    Quantum Optical Hamiltonians}, Phys. Rev. Lett. {\bf 80}, 5465
  (1998).
\bibitem{darlop} G. M. D'Ariano and P. Lo Presti, {\em Quantum
    Tomography for Measuring Experimentally the Matrix Elements of an
    Arbitrary Quantum Operation}, Phys. Rev. Lett.  {\bf 86}, 4195
  (2001).
\bibitem{leung} D. Leung, Ph.D. thesis, Stanford University,  comp-sci/0012017.
\bibitem{sacchialpha} M. F. Sacchi, {\em Maximum-likelihood
    reconstruction of completely positive maps}, Phys. Rev. A {\bf
    63}, 054104 (2001).
\bibitem{faith} G. D'Ariano and P. Lo Presti, {\em Imprinting a
    complete information about a quantum channel on its output state},
  Phys. Rev. Lett. {\bf 91}, 047902 (2003).
\bibitem{altep} J. B. Altepeter, D. Branning, E. Jeffrey, T. C. Wei,
  P. G. Kwiat, R. T. Thew, J. L. O'Brien, M. A. Nielsen, and A. G.
  White, {\em Ancilla-Assisted Quantum Process Tomography}, Phys. Rev.
  Lett. {\bf 90}, 193601 (2003).
\bibitem{calipovm} G. M. D'Ariano, P. Lo Presti, and L. Maccone, {\em
    Quantum Calibration of Measurement Instrumentation}, Phys.  Rev.
  Lett. {\bf 93}, 250407 (2004).
\bibitem{tomo-rew-spring-lopresti} G. M. D'Ariano and P. Lo Presti,
  {\em Characterization of Quantum Devices}, Lect. Notes Phys. {\bf
    649}, 297-322 (Springer, Berlin-New York 2004).
\bibitem{tomo-rew-spring-sacchi} G. M. D'Ariano, M. G. A. Paris, and
  M. F. Sacchi, {\em Quantum Tomographic Methods}, Lect. Notes Phys.
  {\bf 649}, 2-58 (Springer, Berlin-New York 2004).
\bibitem{fiurahrad} J. Fiur\'a\v sek and Z. Hradil, {\em
    Maximum-likelihood estimation of quantum processes}, Phys. Rev. A
  {\bf 63}, 020101(R) (2001).
\bibitem{fiura} J. Fiur\'a\v sek, {\em Maximum-likelihood estimation
    of quantum measurement}, Phys. Rev. A {\bf 64} 024102 (2001).
\bibitem{sanso} A. Luis and L. L. Sanchez-Soto, {\em Complete
    Characterization of Arbitrary Quantum Measurement Processes},
  Phys. Rev. Lett. {\bf 83}, 3573 (1999).
\bibitem{tomo_holography} G. M. D'Ariano, {\em Universal quantum
    observables}, Phys. Lett. A {\bf 300}, 1 (2002).
\bibitem{ccl} A. M. Childs, I. L. Chuang, D. W. Leung, {\em
    Realization of quantum process tomography in NMR}, Phys. Rev. A
  {\bf 64}, 012314 (2001).
\bibitem{dema} F. de Martini, A. Mazzei, M. Ricci, G. M. D'Ariano,
  {\em Exploiting quantum parallelism of entanglement for a complete
    experimental quantum characterization of a single-qubit device},
  Phys. Rev. A {\bf 67}, 062307 (2003).
\bibitem{mohlid} M. Mohseni and D. A. Lidar, {\em Direct
    Characterization of Quantum Dynamics}, Phys. Rev. Lett. {\bf 97},
  170501 (2006).
\bibitem{mohlid2} M. Mohseni and D. A. Lidar, {\em Direct
    characterization of quantum dynamics: General theory}, Phys. Rev.
  A {\bf 75}, 062331 (2007).
\bibitem{mohlid3} M. Mohseni, A. T. Rezakhani, and D. A. Lidar, {\em
    Quantum-process tomography: Resource analysis of different
    strategies}, Phys.  Rev. A {\bf 77}, 032322 (2008).
\bibitem{paz} A. Bendersky, F. Pastawski, J. P. Paz, {\em Selective
    and Efficient Estimation of Parameters for Quantum Process
    Tomography}, Phys. Rev. Lett.  {\bf 100}, 190403 (2008).
\bibitem{delsarte} P. Delsarte, J. M. Goethals, and J. J. Seidel, {\em
    Spherical codes and designs}, Geometriae Dedicata {\bf 6}, 363
  (1977).
\bibitem{lvov} M. Lobino, D. Korystov, C. Kupchak, E. Figueroa, B. C.
  Sanders, and A. I. Lvovsky, {\em Complete Characterization of
    Quantum-Optical Processes}, Science {\bf 322}, 563 (2008).
\bibitem{plenio} J. S. Lundeen, A. Feito, H. Coldenstrodt-Ronge, K. L.
  Pregnell, C. Silberhorn, T. C. Ralph, J. Eisert, M. B. Plenio, I. A.
  Walmsley, {\em Tomography of quantum detectors}, Nature Physics {\bf
    5}, 27 (2008).
\bibitem{opla} S. H. Myrskog, J. K. Fox, M. W. Mitchell, and A. M.
  Steinberg, {\em Quantum process tomography on vibrational states of
    atoms in an optical lattice}, Phys. Rev. A {\bf 72}, 013615
  (2005).
\bibitem{blatt} M. Riebe, M. Chwalla, J. Benhelm, H. H\"affner, W.
  H\"ansel, C. F. Roos, and R. Blatt, {\em Quantum teleportation with
    atoms: quantum process tomography}, New. J. Phys. {\bf 9}, 211
  (2007).
\bibitem{nmr} H. Kampermann and W. S. Veeman, {\em Characterization of
    quantum algorithms by quantum process tomography using quadrupolar
    spins in solid-state nuclear magnetic resonance}, J. Chem. Phys.
  {\bf 122}, 214108 (2005).
\bibitem{solstate} M. Howard, J. Twamley, C. Wittmann, T. Gaebel, F.
  Jelezko, and J. Wrachtrup, {\em Quantum process tomography and
    Linblad estimation of a solid-state qubit}, New J. Phys. {\bf 8},
  33 (2006).
\bibitem{cv} M. Brune, J. Bernu, C. Guerlin, S. Del\'eglise, C.
  Sayrin, S. Gleyzes, S. Kuhr, I. Dotsenko, J. M. Raimond, and S.
  Haroche, {\em Process Tomography of Field Damping and Measurement of
    Fock State Lifetimes by Quantum Nondemolition Photon Counting in a
    Cavity}, Phys. Rev. Lett. {\bf 101}, 240402 (2008).
\bibitem{shor} P. Shor, {\em Scheme for reducing decoherence in
    quantum computer memory}, Phys. Rev. A {\bf 52}, R2493 (1995).
\bibitem{steane} A. M. Steane, {\em Error Correcting Codes in Quantum
    Theory}, Phys. Rev. Lett. {\bf 77}, 793 (1996).
\bibitem{knillaf} E. Knill and R. Laflamme, {\em Theory of quantum
    error-correcting codes}, Phys. Rev. A {\bf 55}, 900 (1997).
\bibitem{mitch} M. W. Mitchell, C. W. Ellenor, R. B. A. Adamson, J. S.
  Lundeen, A. M. Steinberg, {\em Quantum process tomography and the
    search for decoherence-free subspaces}, in {\em Quantum
    Information and Computation II}.  E. Donkor, A. R. Pirich, R.
  Andrew, H. E.  Brandt eds., Proceedings of the SPIE, {\bf 5436},
  223-231 (2004).
\bibitem{emerson} J. Emerson, M. Silva, O. Moussa, C. Ryan, M.
  Laforest, J. Baugh, D. G. Cory, and R. Laflamme, {\em Symmetrized
    Characterization of Noisy Quantum Processes}, Science {\bf 317},
  1893 (2007)
\bibitem{kosut} R. L. Kosut, {\em Quantum Process Tomography via
    L1-norm Minimization}, arXiv:0812.4323.
\bibitem{scott} A. J. Scott, {\em Tight informationally complete
    quantum measurements}, J. Phys. A {\bf 39}, 13507 (2006).
\bibitem{helstrom} C. W. Helstrom, {\em Quantum detection and
    estimation theory}, (Academic Press, New York, San Francisco,
  London, 1976).
\bibitem{mem} G. Chiribella, G. M. D'Ariano, and P. Perinotti, {\em
    Memory Effects in Quantum Channel Discrimination}, Phys.  Rev.
  Lett. {\bf 101}, 180501 (2008).
\bibitem{zim} M. Ziman, {\em Incomplete quantum process tomography and
    principle of maximal entropy}, Phys. Rev. A {\bf 78}, 032118
  (2008).
\bibitem{bapat} R. B. Bhapat, {\em Linear Algebra and Linear Models},
 (Springer-Verlag, New York, 2000).
\bibitem{duffschaeff} R. J. Duffin and A. C. Schaeffer, \emph{A class of nonharmonic Fourier
    series}, Trans. Am. Math. Soc. {\bf 72}, 341 (1952).
\bibitem{li} S. Li, \emph{On general frame decompositions}, Numer. Funct. Anal. Optim. {\bf 16} 1181 (1995).
\bibitem{renormtomo} G. M. D'Ariano, M. F. Sacchi, {\em
   Renormalized quantum tomography} arXiv:0901.2866
\bibitem{magnani} G. M. D'Ariano, D. F. Magnani, and P. Perinotti,
 {\em Adaptive Bayesian and frequentist data processing for quantum
   tomography} Phys. Lett. A, doi:10.1016/j.physleta.2009.01.055.
\bibitem{chl} P. G. Casazza, D. Han, and D. Larson, {\em Frames for
   Banach spaces}, Contemp. Math. {\bf 247}, 149-182 (1999).
\bibitem{defi} C. M. Caves, C. A. Fuchs, and R. Schack, \emph{Unknown quantum states: The quantum de Finetti representation}, J. Math. Phys.
 {\bf 43}, 4537 (2002).
\bibitem{joint} G. M. D'Ariano, P. Perinotti, and M. F. Sacchi,
 \emph{Quantum indirect estimation theory and joint estimates of all
   moments of two incompatible observables}, Phys.  Rev. A {\bf 77},
 052108 (2008).
\bibitem{infolocvsglob} G. M. D'Ariano, P. Perinotti, and M. F.
 Sacchi, \emph{Informationally complete measurements on bipartite
   quantum systems: Comparing local with global measurements}, Phys.
 Rev. A {\bf 72}, 042108 (2005).
\bibitem{spintomo} G. M. D'Ariano, L. Maccone, and M. Paini, \emph{Spin tomography}, J. Opt. B
 {\bf 5}, 77 (2003).
\bibitem{tomosu1}  G. M. D'Ariano, E. De Vito, and L. Maccone, \emph{SU (1,1) tomography}, Phys.
 Rev. A {\bf 64}, 033805 (2001).
\bibitem{su11} G. Chiribella, G. M. D'Ariano, and P. Perinotti, \emph{Applications of the group SU (1,1) for quantum computation and tomography}, Laser
 Phys. {\bf 16}, 1572 (2006).
\bibitem{gromopa} A. Grossmann, J. Morlet, and T. Paul,
 \emph{Transforms associated to square integrable group
   representations. I: General results.}, J. Math. Phys.  {\bf 26},
 2473 (1985).
\bibitem{firstopt} G. M. D'Ariano, P. Perinotti, and M. F. Sacchi,
 {\em Optimization of Quantum Universal Detectors}, in {\em Squeezed
   States and Uncertainty Relations}, ed. by H. Moya-Cessa, R.
 Jauregui, S. Hacyan, and O. Castanos, (Rinton Press, Princeton,
 2003) pag. 86.
\bibitem{tomo_lecture} G. M. D'Ariano, {\em Tomographic methods for
   universal estimation in quantum optics}, Scuola ``E. Fermi'' on
 {\em Experimental Quantum Computation and Information}, F. De
 Martini and C. Monroe ed.  (IOS Press, Amsterdam, 2002) pag. 385.
\bibitem{Gausstomo} G. M. D'Ariano, and N. Sterpi, {\em Robustness
   of Homodyne Tomography to Phase-Insensitive Noise}, J. Mod. Optics
 {\bf 44} 2227 (1997).
\bibitem{proceqcm} G. M. D'Ariano and P. Perinotti, {\em Optimal
   estimation of ensemble averages from a quantum measurement}, in
 {\em Proceedings of the 8th Int. Conf. on Quantum Communication,
   Measurement and Computing}, ed. by O. Hirota, J. H. Shapiro and M.
 Sasaki (NICT press, Japan, 2007), p. 327.
\bibitem{smaps} G. Chiribella, G. M. D'Ariano, and P. Perinotti, \emph{Transforming quantum operations: quantum supermaps},
 Europhys. Lett. {\bf  83}, 30004  (2008).
\bibitem{holevo} A. S. Holevo, \emph{Probabilistic and Statistical
   Aspects of Quantum Theory}, North Holland, Amsterdam, 1982.
\bibitem{scott2} A. J. Scott, \emph{Optimizing quantum process
   tomography with unitary 2-design}, J. Phys. A \textbf{41}, 055308
 (2008).
\bibitem{zeilinger} P. Walther, A. Zeilinger, \emph{Experimental
   Realization of a Photonic Bell-State Analyzer}, Phys. Rev. A {\bf
   72}, 010302(R) (2005)
\bibitem{contmeasure} G. Chiribella, G. M. D'Ariano, D. M.
 Schlingemann, \emph{How continuous quantum measurements in finite
   dimension are actually discrete}, Phys. Rev. Lett. {\bf 98},
 190403 (2007).






\end{thebibliography}
\end{document}